\documentclass[twocolumn]{aastex631}
\received{April 30, 2023}

\submitjournal{ApJ}
\accepted{August 20, 2023}

\shorttitle{Figure Rotation of IllustrisTNG Halos}
\shortauthors{Ash \& Valluri}
\graphicspath{{./}{figures/}}

\begin{document}

\title{Figure Rotation of IllustrisTNG Halos}

\author[0009-0003-7613-3109]{Neil Ash}
\affiliation{University of Michigan Department of Astronomy \\
1085 S. University \\
Ann Arbor, MI 48109, USA}

\author[0000-0002-6257-2341]{Monica Valluri}
\affiliation{University of Michigan Department of Astronomy \\
1085 S. University \\
Ann Arbor, MI 48109, USA}






\begin{abstract}
We use the TNG50 and TNG50 dark matter (DM)-only simulations from the IllustrisTNG simulation suite to conduct an updated survey of halo figure rotation in the presence of baryons. We develop a novel methodology to detect coherent figure rotation about an arbitrary axis and for arbitrary durations and apply it to a catalog of 1,577 DM halos from the DM-only run and 1,396 DM halos from the DM+baryons (DM+B) run that are free of major mergers. Figure rotation was detected in $94\%$ of DM-only halos and $82\%$ of the DM+B halos. The pattern speeds of rotations lasting $\gtrsim 1h^{-1}$ Gyr were log-normally distributed with medians of $0.25~h$ km s$^{-1}$ kpc$^{-1}$ for DM-only in agreement with past results, but $14\%$ higher at $0.29~h$ km s$^{-1}$ kpc$^{-1}$ in the DM+B halos. We find that rotation axes are typically aligned with the halo minor or major axis, in $57\%$ of DM-only halos and in $62\%$ of DM+B halos. The remaining rotation axes were not strongly aligned with any principal axis but typically lay in the plane containing the halo minor and major axes. Longer-lived rotations showed greater alignment with the halo minor axis in both simulations. Our results show that in the presence of baryons, figure rotation is marginally less common, shorter-lived, faster, and better aligned with the minor axis than in DM-only halos. This updated understanding will be consequential for future efforts to constrain figure rotation in the Milky Way dark halo using the morphology and kinematics of tidal streams.
\end{abstract}




\section{Introduction} \label{sec:intro}

Cosmological simulations have predicted that dark matter (DM) halos form in the over-dense regions where large-scale filamentary structures intersect. The consequences of the large-scale structure are two-fold: (1) collapsing protohalos may be torqued  by the local tidal field and (2) matter is accreted onto the halo asymmetrically via a combination of steady accretion and mergers. Both phenomena contribute aspherical density distributions and angular momentum to the forming halo which persist through collapse and virialization \citep{schafer_galactic_2009}. Combined, these two effects may result in figure rotation, a tumbling motion analogous to the rotation of stellar bars in spiral galaxies. Figure rotation is distinct from the possible streaming motions of DM particles within the halo since the latter and does not result in time evolution of a halo's spatial orientation. The commonly used dimensionless halo spin parameter $\lambda$ \citep{peebles_origin_1969, bullock_universal_2001} includes all the angular momentum content of a halo and therefore does not distinguish between streaming motions and figure rotation.

The first measurements of dark halo figure rotation were performed by \cite{dubinski_cosmological_1992}, who performed simulations of 14 isolated halos growing under the influence of a tidal field and observed figure rotation in all halos with pattern speeds between $0.2 - 2.3 h$ km s$^{-1}$ kpc$^{-1}$ over a $\sim 1 h^{-1}$ Gyr period. \cite{bailin_figure_2004} conducted a survey of figure rotation in an N-body DM-only $\Lambda$CDM cosmological simulation with a box of side length 50 $h^{-1}$ Mpc and particle mass of $\sim 8\times10^7 ~h^{-1} ~M_\odot$. Over the surveyed $1~h^{-1}$ Gyr duration, they find that upwards of 90$\%$ of halos within their selected sample undergo coherent figure rotation, with $\sim 85\%$ rotating about the minor axis and $\sim 15\%$ about the major axis. The pattern speeds for these rotations were log-normally distributed with a mean of 0.15 $h$ km s$^{-1}$ kpc$^{-1}$  (9$^\circ ~h$ Gyr$^{-1}$) and width of 0.83 $h$ km s$^{-1}$ kpc$^{-1}$. A later study from \cite{bryan_figure_2007} expanded on this work by searching for coherent rotations over a 5 $~h^{-1}$ Gyr period, and found that over this period only 5 of their 222 halo sample ($\gtrsim 2\% $) underwent steady rotation. For the last $1~h^{-1}$ Gyr, the observed fraction of halos undergoing figure rotation rose to $61\%$, with mean pattern speeds similar to those found by \cite{bailin_figure_2004}. Their measured pattern speeds showed a systematic decrease with duration; halos which rotated steadily over $1~h^{-1}$ Gyr had a mean pattern speed similar to that found by \cite{bailin_figure_2004} of $\sim 0.24 h$ km s$^{-1}$ kpc$^{-1}$, whereas for halos rotating over 5 $~h^{-1}$ Gyr this mean dropped to only $0.09 h$ km s$^{-1}$ kpc$^{-1}$. \cite{bryan_figure_2007} also found a much weaker alignment between the rotation axis and the halo minor axis. These axes were aligned in only half their halos. Apart from the 1, 3, and 5 $~h^{-1}$ Gyr bins studied by \cite{bryan_figure_2007}, there has been no detailed investigation of the time evolution and longevity of figure rotation. 

In order to perform future measurements of figure rotation in dark halos, we first need a more developed understanding of the parameters describing the figure rotation (e.g. the rotation axis orientations, pattern speeds, rotation durations) that we are likely to observe in galaxies in the presence of baryons. Previous studies of figure rotation have been limited to DM-only $\Lambda$CDM simulations. However, the influence of baryonic physics could be important for figure rotation. It is has been well established that the baryons in the inner halo can transform the shapes of inner regions of dark matter halos from triaxial to oblate \citep[e.g.][]{kazantzidis_effect_2004, zemp_impact_2012,chua_shape_2019, prada_dark_2019} and can exchange angular momentum with the dark halo 
\citep{duffy_impact_2010,bryan_impact_2013}. Whether angular momentum transferred from the baryonic component to the dark halo can influence figure rotation remains unknown. To our knowledge a study of the behavior of figure rotation in the presence of baryonic physics has not been carried out to date.

In this paper we conduct a study of relatively steady-state cosmological halos to understand the time evolution and stability of figure rotation in steady state and in the presence of baryons. The objective of this work is to quantify the influence of baryonic physics on dark halo figure rotation during secular evolution. To this end, we follow the examples of \cite{bailin_figure_2004} and \cite{bryan_figure_2007} of identifying a catalog of merger-free ``quiescent'' halos in the TNG50 run of the IllustrisTNG simulation suite \citep{nelson_illustristng_2018,nelson_first_2019,pillepich_first_2019}. We take advantage of the complementary DM-only and full DM+baryon (DM+B) runs to directly track the influence of baryons in halos matched between DM-only and DM+B runs \citep{nelson_illustris_2015}. We make measurements of figure rotation and the durations of these rotations over a $2.7h^{-1}$Gyr $\sim 4$ Gyr time course ($ 0 \lesssim z \lesssim 0.35$) using novel methods which can sensitively detect both the orientation of rotation axis independent of its alignment or misalignment with the halo principal axes and the duration of figure rotation. The results of this study will be instrumental in future efforts to detect dark halo figure rotation.

The remainder of this paper is outlined as follows. In section \ref{sec:simulations} we introduce the TNG50 simulation suite utilized for this work. In section \ref{sec:methods} we describe the methodology developed and utilized within this study. Section \ref{sec:results} presents our key results and findings applying our methodology to TNG50, and we conclude with some discussion of these results in section \ref{sec:conclusion} and future directions.

We present all numbers and results scaled as $h$-independent units for ease of comparison with past results. We make use of the Planck2015 cosmology \citep{planck_collaboration_planck_2016} where $h = 0.6774$.

\section{Simulations}\label{sec:simulations}

\subsection{IllustrisTNG}

For this work we use the TNG50 simulation suite \citep{nelson_illustristng_2021,nelson_first_2019, pillepich_first_2019}. TNG50 is the highest resolution realization of the IllustrisTNG cosmological simulation suite. The simulation is run using the cosmological magnetohydrodynamical moving-mesh simulation code \texttt{AREPO} \citep{weinberger_arepo_2020} in a simulation box of side length 50 comoving Mpc with $2\times2160^3$ DM and gas particles, giving a DM mass resolution of $4.5\times 10^5$ M$_\odot$ and spatial resolution of roughly 100-140 pc. TNG50 was run using the Planck2015 cosmology \citep{planck_collaboration_planck_2016} with H$_0 = 67.8$ km s$^{-1}$ Mpc$^{-1}$, $\Omega_m = 0.308$, and $\Omega_bh^2 = 0.022$. We make use of the final 25 snapshots, spanning redshifts of $ 0 \lesssim z \lesssim 0.35$. Within this range, the mean temporal resolution is $107 h^{-1}$ Myr, with a minimum of $80 h^{-1}$ Myr and maximum of $160 h^{-1}$ Myr. 

Halo and subhalo catalogs for TNG50 are generated using the FOF and subfind algorithms, respectively. Halos are tracked across snapshots by the LHaloTree merger tree, generated using the FOF halos \citep{springel_cosmological_2005}.
We make use of the bi-direction DM-only/ DM+B subhalo matching catalog generated using LHaloTree \citep{nelson_illustris_2015}.

\subsection{Halo catalog}\label{sec:halo_catalog}

\begin{figure}
    \centering
    \includegraphics[width=0.4\textwidth]{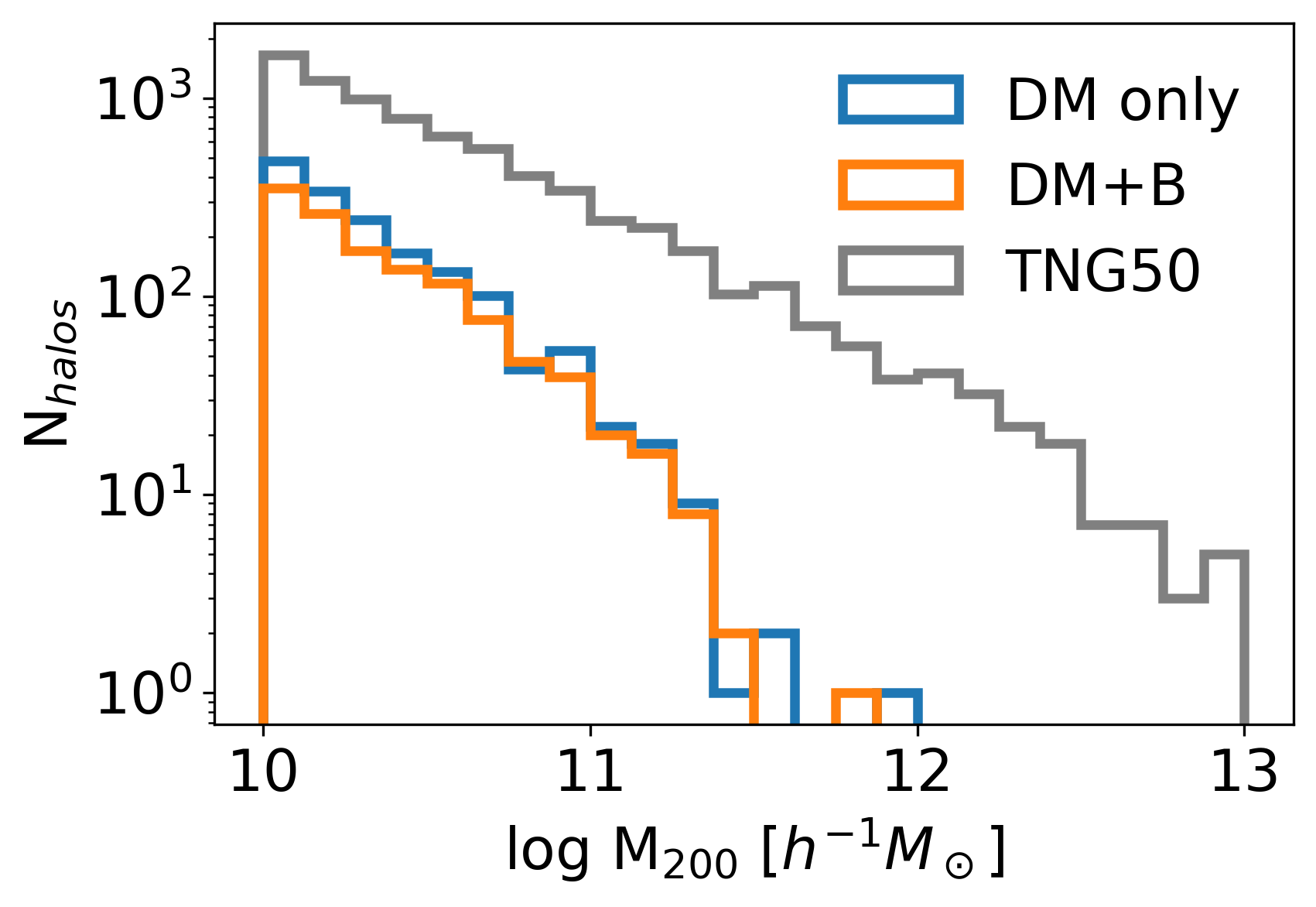}
    \caption{Halo mass function of the selected halo catalog for the DM only (blue) and DM+baryon (orange) TNG50 runs compared to the full DM+B TNG50 simulation mass function. Our selection criteria omit halos with large mass accretion rates, thereby preferentially selecting lower mass halos.}
    \label{fig:mass_selection_effects}
\end{figure}

We produce a catalog of halos in the mass range $10^{10}- 10^{13} ~h^{-1}~M_\odot$ which are free of massive mergers for the last $\sim 3~h^{-1}$ Gyr in order to probe the behavior of figure rotation in quiescence, without significant gravitational perturbations from outside sources. Within the sampled time course, major-merger free halos are selected according to the following criteria:
\begin{enumerate}
    \item \label{catalog_1} The mass accretion rate $\dot{M}_{200}$ never exceeds 10$\%$ of the virial mass $ ~M_{200}$ in 2 snapshots ($\sim 200 ~h^{-1}$ Myr)
    \item \label{catalog_2} The halo at $z=0$ has a defined progenitor within the halo tree at each snapshot. This may not be the case if the halo finder cannot identify a clear progenitor in one snapshot, and additionally protects against flyby interactions.
    \item \label{catalog_3} The cumulative mass in all subhalos does not exceed 5$\%$ that in the host's main subhalo at any point in the sampled time course
\end{enumerate}
Each of these criteria serve to eliminate halos that undergo a major merger (defined as a merger whose secondary:primary mass ratio $f \geq 0.10$) or have massive companions. It is important to remove these halos, as orbiting mass from a satellite can bias our shape tensor metric and can produce an artificial figure rotation signature. We find these criteria to be fairly restrictive; of the 10,645 halos within the specified mass range, 1,910 halos satisfy items \ref{catalog_1}-\ref{catalog_3} within the time course, or roughly $18\%$. Our results from this population are therefore not representative of the global halo population, but rather a population of quiescent, merger-free halos. These criteria are similar to those used by \cite{bailin_figure_2004}, who remove halos with satellites above $\sim 5\%$ the total halo mass. It should be noted that these criteria explicitly exclude MW-LMC like systems, and preferentially select lower-mass halos (see figure \ref{fig:mass_selection_effects}).  

The catalog of halos for the DM+B runs was generated using the LHaloTree matching catalog introduced in \cite{nelson_illustris_2015}. This catalogue matches subhalos with bi-directionality for the DM-only and DM+B runs. We obtain the DM+B analogs for each DM-only halo in our catalog by matching the main subhalo to the corresponding main subhalo in the baryon runs. For 15 of the DM-only subhalos, the matching subhalo in the DM+B simulation was not the main subhalo of its host group. We omit these halos from the baryon catalog, as we are interested only in the main halo and not satellite-like subhalos. Only one subhalo in our DM-only catalog did not have a matching DM+B subhalo. With these exceptions, our DM+B catalog was reduced by 16 halos compared to our DM-only catalog. This reduced the number of halos by $\lesssim 1\%$. We then explicitly check that each of the halo analogs from the DM+B simulation follow criteria \ref{catalog_1} - \ref{catalog_3}.  Applying these criteria removed another 140 halos from the original catalog ($\sim 7\%$). Using the outlined halo selections above, we arrive at a final catalog of 1,910 DM-only halos and 1,754 DM+B halos.

\section{Methods} \label{sec:methods}

\subsection{Shape determination}

We determine the shapes and orientations of the halo principal axes using the standard procedure iterating over the shape tensor  \citep[see e.g.][]{emami_morphological_2021}. In summary, we select particles within the ellipsoidal volume $0 \leq |r_\mathrm{ell}| < 0.6R_{200}$, where the ellipsoidal radius $r_\mathrm{ell}$ is given as
\begin{equation}
    |r_\mathrm{ell}|^2 = \left(\frac{r_{{\rm body},1}}{a}\right)^2 + \left(\frac{r_{{\rm body},2}}{b}\right)^2 + \left(\frac{r_{{\rm body},3}}{c}\right)^2
    \nonumber
\end{equation}
and 
\begin{equation}
    \vec{r}_\mathrm{body} = \mathbf{E}^{\text{T}}(\vec{r} - \vec{r}_0).
\end{equation}
$\vec{r}_\mathrm{body}$ describes particle positions within the halo "body" frame (i.e., its position with respect to the halo principal axes). The halo center $\vec{r}_0$ is taken to be the position of the most bound halo particle (i.e., where the potential is lowest). $a$, $b$, and $c$ are the halo shape parameters and $\mathbf{E}$ gives the directions of the principal axes. We initialize these values with $a=b=c=1$ and $\mathbf{E} = \mathbf{I}$, and update the values by diagonalizing the shape tensor whose elements are 
\begin{equation}
    S_{i,j} = \frac{1}{M}\sum_{k=1}^N m_k (r_k)^i (r_k)^j.
\end{equation}
The axis lengths $a$, $b$, and $c$ are proportional to the square root of the eigenvalues of the shape tensor, and are normalized such that the ellipsoidal shell maintains a constant volume during iteration. The eigenvectors define the orientation of the halo principal axes $\mathbf{E}$, and are updated each iteration. Iteration is terminated once the cumulative difference in the normalized eigenvalues between iterations is below 0.01:
\begin{equation}
    \left|\left(\sum_{i=1}^3\frac{\lambda_i}{(\lambda_1\lambda_2\lambda_3)^{1/3}}\right)^j - \left(\sum_{i=1}^3\frac{\lambda_i}{(\lambda_1\lambda_2\lambda_3)^{1/3}}\right)^{j-1}\right| < 0.01,
\end{equation}
where $\lambda_i$ is a given eigenvalue, and $j$ is the iteration number. Note that each $\lambda_i$ is strictly positive.

The shape tensor's eigenvectors are degenerate with reflections about the origin (that is, if $\vec{r}$ is an eigenvector of shape tensor $\mathbf{S}$, then so is $-\vec{r}$). This parity degeneracy creates a challenge for measuring the evolution of an axis across snapshots. We choose the axis parity which minimizes the angle between a given axis in subsequent snapshots, subject to the constraint that both sets of axes make up a right-handed basis set (i.e. $|\mathbf{E}| = 1$ for both snapshots). If the minimum angle between any principal axis in subsequent snapshots exceeds 90$^\circ$, then we discard that halo. We justify this cut on the basis that a pattern speed $\Omega_p \gtrsim 90^{\circ} / \text{snapshot} \sim 14~h$kms$^{-1}$kpc$^{-1}$ is much faster than those observed by previous studies \citep{bryan_figure_2007, bailin_figure_2004, dubinski_cosmological_1992} and lies well above the upper stability limit of $\sim 1.8~ h $ km s$^{-1}$ kpc$^{-1}$ for orbits maintaining triaxiality \citep{deibel_orbital_2011}. For the typical snapshot spacing in IllustrisTNG we are therefore unlikely to observe halos with pattern speeds greater than the Nyquist frequency. \cite{bailin_figure_2004} perform a similar procedure by reflecting major axes through the origin until the angle between major axes of subsequent snapshots is less than 90 degrees. Only 1 of the 1,910 halos in the DM only catalog and 2 of the 1,754 halos in the DM + baryons catalog were removed by this criterion. 

\subsection{Axis orientation uncertainty}\label{sec:error}

\begin{figure*}
    \centering
    \includegraphics[width = \textwidth]{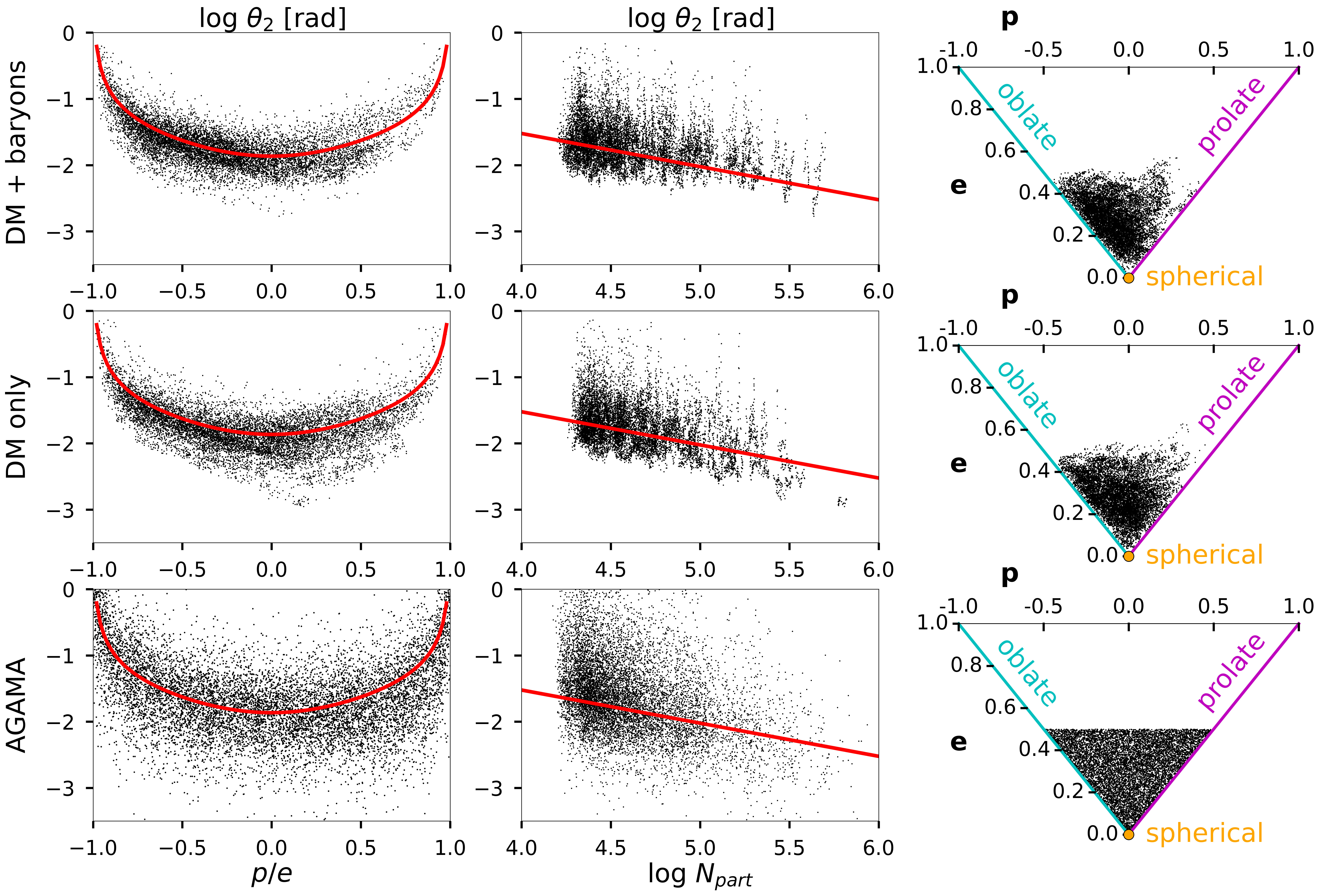}
    \caption{Intermediate axis orientation uncertainties plotted against halo shape (left column) and number of particles (middle column) for a subsample of 10,000 random TNG50 DM-only, TNG50 with baryons, and randomly generated AGAMA halos. The right column shows the distributions of halos in the $e$-$p$ shape parameter space, which are bounded by the triangle with sides $e=1$ (fully flat halo), $p=e$ (fully prolate halo), $p=-e$ (fully oblate halo). Overplotted in the left and middle columns is the analytic error prescription defined in equation \ref{eq:analytic error}, assuming an average number of halo particles (left column) and average halo shape (middle column). The errors for the DM-only and DM+B TNG50 runs appear functionally similar, while errors on the AGAMA halos \replaced{are $\sim 1$ dex smaller but show a similar dependence on halo shape}{show a similar overall trend with larger scatter.} Errors for both TNG50 runs are determined using the jackknife sampling routine described in section \ref{sec:error}.}
    \label{fig:analytic_error_fits}
\end{figure*}

The accuracy to which we can determine the orientation of a halo's principal axes is dependent on both the halo shape and the number of particles it contains \citep{bailin_figure_2004}. The effects of shape can be understood intuitively. Prolate (1 long axis and 2 short axes), oblate (2 long axes and 1 short axis), and spherical (all axes equal) halos each have at least 2 axes whose orientations are undefined because they have equal lengths. In such cases, the uncertainty on the axis orientation effectively diverges. Conversely, in a triaxial halo each of the 3 axes have different lengths and can be accurately identified, giving the minimum orientation uncertainty. Individual particle positions are affected by Poisson noise, and hence the orientation uncertainty will also scale inversely with $\sqrt{N}$ (see the middle column in figure \ref{fig:analytic_error_fits}). 

To quantify halo shape, we use modified ellipticity and prolateness metrics, defined as
\begin{equation}
    e = 1 - \frac{c}{a}; \quad p = 1 - 2\frac{b}{a} + \frac{c}{a},
\end{equation}
where $c/a$ and $b/a$ are the axis ratios of the minor:major and intermediate:major axis lengths, respectively. Unlike the common definition used by e.g. \cite{despali_like_2014} and \cite{dubinski_cosmological_1992}, in this form all halos are bound to the triangle defined by  $-e \leq p \leq e$, $e \leq 1$. The ratio $p/e$ allows us to parameterize halo shape using a single variable. $p/e = -1$ corresponds to an oblate halo, and $p/e = 1$ indicates a prolate halo. We find that typically TNG50 halos are bound within this triangle by $e \lesssim 0.5$ (see figure \ref{fig:analytic_error_fits}.)

For both the DM only and DM+B TNG50 runs, we select a random subset of 400 halos taken from the catalog identified in section \ref{sec:halo_catalog} to quantify the functional dependence of the orientation uncertainty on halo shape $p/e$ and particle number $N_{\mathrm{part}}$. For every halo within the subset, we generate $100$ mock halo halos by randomly sampling $90\%$ of the halo particles without replacement. This method is analogous to a jackknife sampling routine. For each realization, we measure the orientation of the three halo principal axes. The uncertainty on the orientation of a given principal axis is estimated using the angular spread in the jackknife sampled axis orientations about the mean axis orientation. Because this method reduces the number of particles used to measure axis orientation, there is added noise in the measurement. This noise is Poisson distributed \citep{bailin_figure_2004}, and is proportional to $1/\sqrt{N_{\mathrm{part}}}$. We therefore multiply the recovered jackknife uncertainty by a factor of $\sqrt{0.9}$ to remove the added noise introduced by our jackknife resampling. We prefer our jackknife sampling routine using $90\%$ of the particles to a bootstrapping routine, as it is significantly less computationally intensive and allows us to more naturally account for the added Poisson noise caused by resampling. The jackknife sampling routine was repeated for the 400 randomly selected halos for each of the 25 snapshots spanning our 2.7 $h^{-1}$ Gyr time course, giving a total of 10,000 uncertainty estimates.

To independently verify the uncertainties observed in our jackknife halos, we compare to a set of 10,000 halos randomly generated using the AGAMA software \citep{vasiliev_agama_2019}. These halos were generated with \added{virial} masses drawn from \replaced{a Press-Schechter halo mass function \citep{press_formation_1974} in the halo catalog mass range of $10^{10} - 10^{13} h^{-1} M_\odot$}{the distribution of halo virial masses in our catalog (see fig. \ref{fig:mass_selection_effects})}. They were generated with NFW density profiles \citep{navarro_universal_1997} and concentrations determined by the mass-concentration relation identified by \cite{bullock_profiles_2001}. The mass of each particle was taken to be identical to the particle mass in the DM-only TNG50 run. We generate these halos with random shapes uniformly covering the $p - e$ parameter space \added{where $e<0.5$}. Each halo is generated with a known but random spatial orientation, which allows us determine the error in our measured orientation without the need for jackknife sampling. 

Using the sample of jackknife halos, we confirm that the orientation uncertainty scales as $1/\sqrt{N}$ (middle column of figure \ref{fig:analytic_error_fits}). The left column of figure \ref{fig:analytic_error_fits} shows the dependence of the orientation uncertainty for the halo intermediate axis against halo shape. We find that for most triaxial shapes ($|p/e| << 1$) the orientation uncertainty is similar for each of the three principal axes, with the minor axis showing slightly lower uncertainties and the intermediate axis slightly higher. For halos which are nearly oblate or prolate ($|p/e| \sim 1$) the cumulative uncertainty on halo orientation becomes large. This behavior is consistent between TNG50 halos and AGAMA halos, though AGAMA halos \deleted{show errors which are typically $\sim 1$ dex lower than TNG50 halos and} show a considerably higher degree of scatter, as can be seen in figure \ref{fig:analytic_error_fits}. \added{While particles in substructure identified by TNG50's subfind algorithm were explicitly removed prior to the jackknife error measurements, it is possible that residual particles originating from and surrounding the removed substructure remain for some halos. These residual particles could influence our shape and orientation measurements. The sampled AGAMA halos were drawn from a $p$ and $e$ parameter space which does not perfectly match the TNG50 halo shape parameter space, producing a larger range in the sampled shape parameter space which could yield a greater scatter at a fixed $p/e$ shape.}

We determine analytical approximations to the axis orientation uncertainties for use during our measurements of figure rotation. The functional forms of the errors on each axis are well-described by hyperbolic cosecant functions:

\begin{equation}
    \begin{array}{cc}
      \theta_1 =&1.01\times10^{-2} \sqrt{N} \text{csch}[8.17\times10^{-1}(1+p/e)],   \\
      &\\
      \theta_2 =&2.11\times10^{-2} \sqrt{N}  \big\{\text{csch}[1.65(1+p/e)] + \\ &~~~~~~~~~~~~~~~~~~~~~~~\text{csch}[2.52(1-p/e)]\big\}, \\
      &\\
       \theta_3 =&5.34\times10^{-6} \sqrt{N} \text{csch}[5.84\times10^{-4}(1-p/e)],
    \end{array}
    \label{eq:analytic error}
\end{equation}
where $\theta_1, \theta_2, \theta_3$ correspond to the orientation uncertainties on the major, intermediate, and minor axes, respectively. The coefficients listed in eq \ref{eq:analytic error} were determined by least-squares fitting on the jackknife halos for the DM-only TNG50 run. The curves for the $\theta_2$ fit is overplotted on fig \ref{fig:analytic_error_fits}. We find that the fit results using the DM-only TNG50 run provide a good description of the errors for the full DM+B TNG50 run, and so we use the same expression and fit coefficients for each TNG50 run. 

We find that the mean $p/e$ shapes in the DM-only and DM+B TNG50 runs are $-0.28$ and $-0.14$ respectively, implying that TNG50 DM halos are more oblate in the presence of baryons, consistent with expectations from previous work  \citep[e.g.][]{chua_shape_2019}. For these values, the anticipated uncertainties on the major axis orientation would be $0.022$ and $0.016$ radians for DM+B and DM-only halos. The uncertainty on axis orientation becomes very large for $|p/e| \gtrsim 0.9$, and so we remove these halos from our analysis. This criterion removes 332 of the 1,909 DM-only halos and 356 of the 1,752 DM+B halos, corresponding to a reduction by $17\%$ and $20\%$, respectively. Thus the remaining number of halos used for measurements of figure rotation were $1,577$ DM-only halos and $1,396$ DM+B halos. Despite our restrictive selection criteria this set offers us much improved population statistics relative to past studies. \cite{bailin_figure_2004} study a population of 317 halos and \cite{bryan_figure_2007} a population of 222 halos. The improvement in sample size is owed to the much higher particle resolution in TNG50, which allows us to expand our search to much lower mass halos than were previously accessible. \cite{bailin_figure_2004} and \cite{bryan_figure_2007} both limit their studies to halos with at least 4000 particles, placing lower mass limits at $\sim 3\times 10^{11} h^{-1}$ and $2.4\times10^{12} h^{-1} M_\odot$, respectively. Adopting a similar constraint would place our lower mass limit at $\sim 1.5 \times 10^9 h^{-1} M_\odot$. We instead choose to truncate our sample at a lower mass limit of $10^{10} h^{-1} M_\odot$, to keep the sample size tractable.

\subsection{Rotation axis fitting}\label{sec:rotation axis fitting}

Past studies of figure rotation in cosmological N-body simulations predominantly relied on a best-fit plane method, first described in \cite{bailin_figure_2004}. In summary, this method first identifies the orientation of the halo major axis at each snapshot and then attempts to fit a plane of the form $z = ax + by$ to the major axes from all snapshots. By projecting the major axes into the best fit plane, the pattern speed can be inferred. This method is robust for rotation axes perpendicular to the major axis, but as noted by \cite{bailin_figure_2004} is insensitive to cases where the rotation axis is either aligned with the major axis or misaligned with all three axes. To mitigate this, they rely on a supplementary quaternion method, which is sensitive to arbitrary rotations but cannot readily combine rotations from multiple snapshots. We seek to generalize the plane method developed by \cite{bailin_figure_2004} to be sensitive to rotations about arbitrary axes. We do this assuming figure rotation resembles a simple rotation about a fixed axis similar to \cite{dubinski_cosmological_1992}. This assumption may not be strictly true if the halo rotation axis is able to precess, but remains a reasonable assumption if the precession angle or frequency are small.

\begin{figure*}
    \includegraphics[width=.5\textwidth]{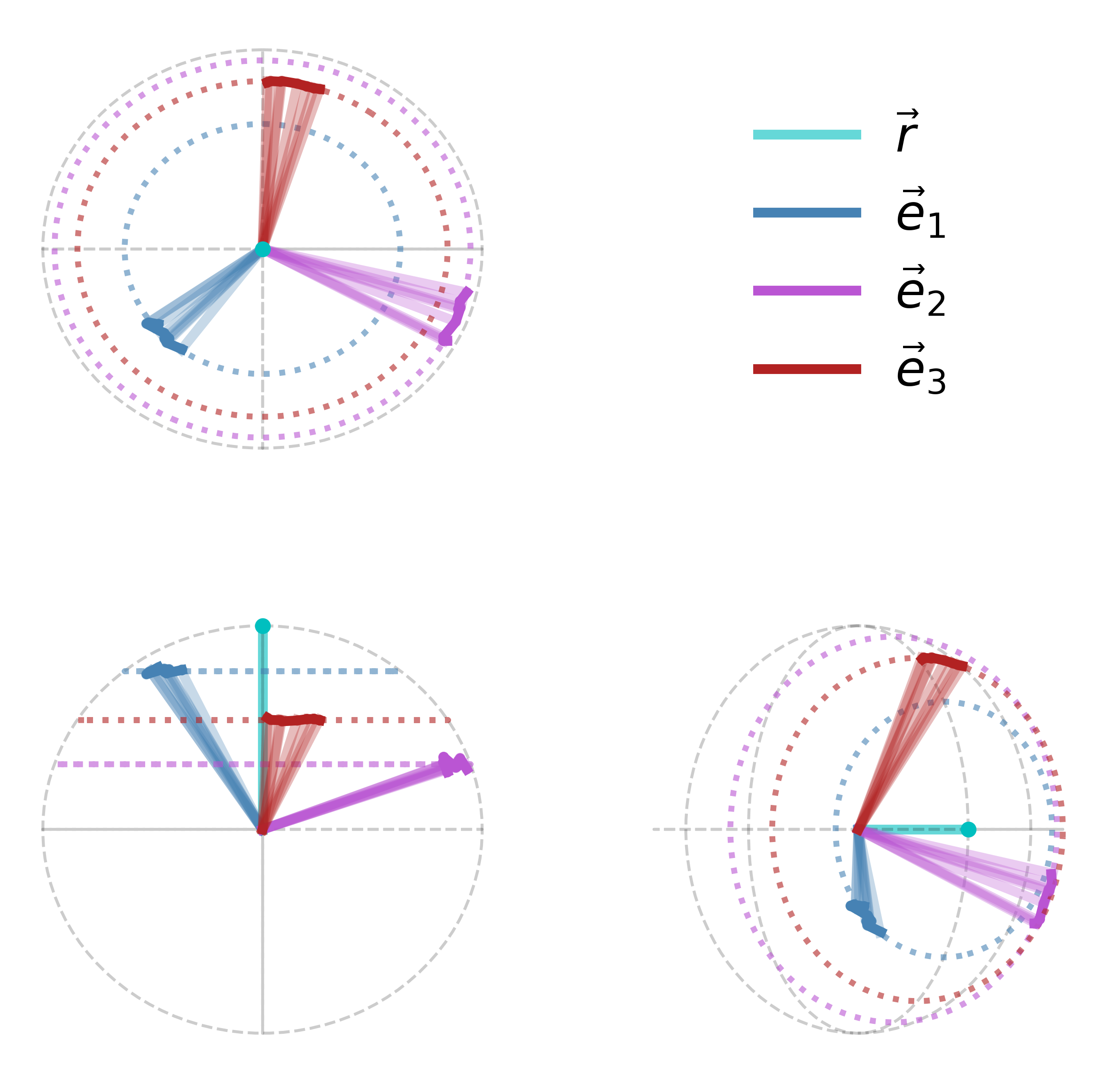}
    \includegraphics[width=.5\textwidth]{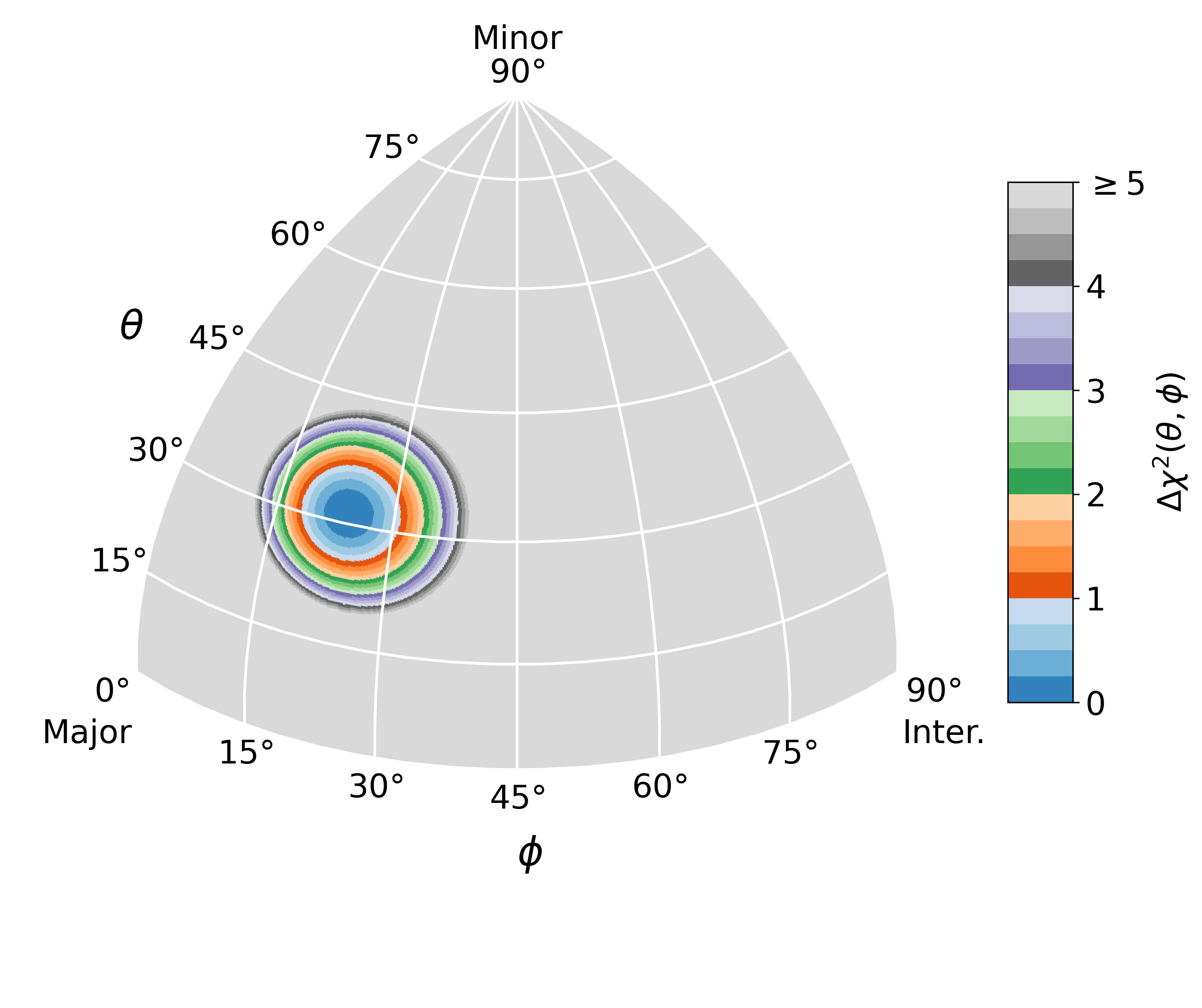}
    \caption{Example of the single-axis fitting process described in section \ref{sec:rotation axis fitting}. The plots to the left show the tracks traced out by each of the three principal axes during a stable rotation period as viewed from a position along the rotation axis shown in cyan (top left), a position perpendicular to the rotation (bottom left) and from an arbitrary viewing angle (bottom middle). The motion of the principal axes are well described by three parallel planes, as discussed in section \ref{sec:rotation axis fitting}. The figure to the right shows the localization of the rotation axis in the halo body frame (represented by an octant showing spherical-polar angles $\theta, \phi$) using equation \ref{eq:single_axis_chi2}. The plot is colored by the change in the fit $\chi^2$ as the rotation axis orientation deviates from the best-fit value. $\Delta\chi^2 = 1$ corresponds approximately to a 1$\sigma$ threshold on the orientation of the rotation axis. In this example, our methodology recovers the orientation of the rotation axis to within $\sim 5^\circ$.}
    \label{fig:single axis fitting}
\end{figure*}

Under a general rotation, the motions of each of the three principal axes will be contained by a plane with some offset from the halo center. These three planes are all mutually parallel, and the offsets of these planes from the halo center give the three components of the rotation axis in the body frame of the halo. We demonstrate this as follows. Let the orientation of a given principal axis be represented as the unit vector in time as $\hat{e}_j(t)$ with $j=1,2,3$ indicating the major, intermediate, and minor axes, respectively, and consider a simple rotation about a single axis oriented along a unit vector $\hat{r}$. Due to the invariance of dot products under rotation, we can write that 
\begin{equation}
    \hat{e}_j(t)\cdot \hat{r} = \text{cos}\theta_j = \text{const},
\end{equation}
Where $\theta_j$ is the angle between $\hat{e}_j(t)$ and $ \hat{r}$. This expression defines a plane perpendicular to $\hat{r}$ and containing the point $\hat{r} \text{cos}\theta_j$. To understand the meaning of these constants, we transform the rotation axis into the body frame according to the change of basis matrix
\begin{equation}
    \mathbf{E}^{\text{T}}\hat{r}\mathbf{E} = [\hat{e}_1,\hat{e}_2,\hat{e}_3]^{\text{T}}\hat{r} [\hat{e}_1,\hat{e}_2,\hat{e}_3] = \left[
\begin{array}{cc}
\text{cos}\theta_1 \hat{e}_1\\ 
\text{cos}\theta_2 \hat{e}_2\\ 
\text{cos}\theta_3 \hat{e}_3
\end{array}     \right].
\end{equation}
Hence, the offsets $\text{cos}\theta_j$ specify the rotation axis within the body frame of the ellipsoid. This property arises purely from the invariance of dot products under rotation and because the three principal axes make up a complete basis set (by definition). The left panels of figure~\ref{fig:single axis fitting} show three projections of the principal axes and rotation axis for a sample TNG halo undergoing figure rotation, for illustration. We note that while general 3D rotations are commonly described using three Euler angles, they can equivalently be characterized by a rotation axis and angle of rotation (i.e., the quaternion representation). We therefore lose no generality in this formulation.

While it is possible to simultaneously fit one plane to each of the three principal axes to determine the axis of rotation, doing so requires fitting 6 parameters (3 components of $\hat{r}$ and $a_0,a_1,a_2$) which are mutually dependent in a non-linear way. These 6 parameters are more simply described as the rotation axis in the simulation box frame ($\hat{r}$) and in the body frame of the ellipsoid ($\hat{a}$) which approximates the halo. The rotation axis can more simply be specified using only 2 parameters (the spherical polar components of the axis $\theta$ and $\phi$). With this in mind, we define a routine to identify the rotation axis in the halo body frame which fits only these 2 parameters. 

Under a simple rotation (i.e., neglecting precession or other time-evolution), the orientation of the rotation axis will remain fixed both in the simulation box frame and within the halo body frame. We show this as follows. The transformation between the two frames is done by 
\begin{equation}
    \hat{r} = \mathbf{E}(t) \hat{a},
    \label{eq:box_frame_to_body_frame_rAxis}
\end{equation}
where $\mathbf{E}(t) = [\hat{e}_1(t), \hat{e}_2(t), \hat{e}_3(t)]$ is the matrix containing the principal axis orientations at a time $t$. This transformation is valid since we have defined $\mathbf{E}(t)$ such that it is always a right-handed basis set. At some later time, the principal axes have a new orientation $\mathbf{E}(t+\Delta t)$. The rotation matrix describing this evolution is 
\begin{equation}
    \mathbf{R} = \mathbf{E}(t+\Delta t)\mathbf{E}^{-1}(t).
    \nonumber
\end{equation}
By definition, the axis $\hat{r}$ about which the halo has rotated does not change under the rotation:
\begin{equation}
    \mathbf{R}\hat{r} = \hat{r}.
    \nonumber
\end{equation}
It then follows that 
\begin{equation}
    \begin{array}{cc}
        \hat{r} =& \mathbf{E}(t+\Delta t)\mathbf{E}^{-1}(t) \hat{r}, \\
        \hat{r} =& \mathbf{E}(t+\Delta t)\mathbf{E}^{-1}(t) \mathbf{E}(t) \hat{a}, \\ &\\
        \hat{r} =& \mathbf{E}(t+\Delta t) \hat{a}.
    \end{array}
    \nonumber
\end{equation}
Hence, equation \ref{eq:box_frame_to_body_frame_rAxis} holds at all times during a steady, coherent rotation, and the rotation axis in both the simulation box frame and in the halo body frame are constant. 

The only vectors which are invariant under a rotation $\mathbf{R}$ are $\hat{r}$ and $-\hat{r}$, implying that $\hat{a}$ and $-\hat{a}$ are the only two vectors in the halo body frame which do \added{not} change orientation in the simulation box frame during the rotation. Because the orientations of the principal axes are uncertain, and because rotations are not perfectly steady, the rotation axis in the box frame will not be perfectly constant, but will take some value $\hat{r}(t)$ at a given snapshot which has some angular deviation from the true rotation vector $\hat{r}$. This angular deviation can be written as 
\begin{equation}
    \Delta \theta(t) =  \text{arccos}(\hat{r}(t) \cdot \hat{r}).
    \label{eq:angular_dev_1}
\end{equation}
In practice, we do not know the true value of $\hat{r}$, and so we take it to be the approximate mean over all the relavant snapshots:
\begin{equation}
    (\hat{r})^j \sim \left<\mathbf{E}(t)\cdot\hat{a}\right>^j = \frac{ 1}{N_{snap}}\sum_{t=t_i}^{t_f} (\mathbf{E}(t)\cdot\hat{a})^j
\end{equation}
Where the superscript $j$ corresponds to the $j-$th component of $\hat{r}$, and $N_{snap}$ is the number of snapshots the average is taken over. The average vector $\left<\mathbf{E}(t)\cdot\hat{a}\right>$ is normalized to unit length after each component is calculated. With this definition, we may rewrite eq \ref{eq:angular_dev_1} as:
\begin{equation}
    \Delta \theta(t) = \text{arccos}[\mathbf{E}(t)\hat{a} \cdot \left<\mathbf{E}(t)\cdot\hat{a}  \right>]
\end{equation}
where we have substituted $\hat{r}(t) = \mathbf{E}(t)\hat{a} $ using equation \ref{eq:box_frame_to_body_frame_rAxis}. The expression allows us to define both the sum of square errors over a time course of snapshots and a reduced $\chi^2$ for some choice of rotation vector $\hat{a}$:
\begin{equation}
    \chi^2_\nu = \frac{1}{\nu} \sum_{t=t_i}^{t_f} \frac{\Delta \theta^2(t)}{\sigma_\theta^2}.
    \label{eq:single_axis_chi2}
\end{equation}
Here, we choose to approximate the degrees of freedom as $\nu = N_{snap} - 2$, where the choice of a mean rotation axis removes 2 degrees of freedom. $\sigma_{\theta}$ gives the expected angular uncertainty on the rotation axis which is predicted by interpolating from the errors on each of the three principal axes, determined as described in section \ref{sec:error}. Note that $\chi^2_\nu$ is undefined for $N_{snap}=2$, because $\nu = 0$. This behavior is expected; a rotation over two snapshots always has a uniquely determined rotation axis, which can be determined by calculating a quaternion between the two snapshots. Because the rotation axis is uniquely determined by the quaternion in this case, the degrees of freedom must be zero for rotations between subsequent snapshots.

To determine the best fit rotation axis to describe the evolution of a halo over a series of snapshots, we minimize the sum of squared errors $\sum \Delta \theta^2$ for a series of snapshots. The overall fit quality is then determined by equation \ref{eq:single_axis_chi2}. Using this methodology, we are able to recover the best-fit rotation axis in a completely general way which does not require the rotation axis to align with any principal axis. This method is robust against halos which are not undergoing figure rotation. Halos whose axes do not evolve in time will be identified during the pattern speed measurement, and halos that undergo random, incoherent rotations will show large $\chi^2_{\nu}$ which identifies them as not undergoing steady figure rotation. Figure \ref{fig:single axis fitting} (right) shows an example localization, with the corresponding motions of the halo principal axes in the simulation box frame.

\subsection{Determination of rotation duration}\label{sec:rotation_epoch_determination}

\begin{figure*}
    \includegraphics[width=0.635\textwidth]{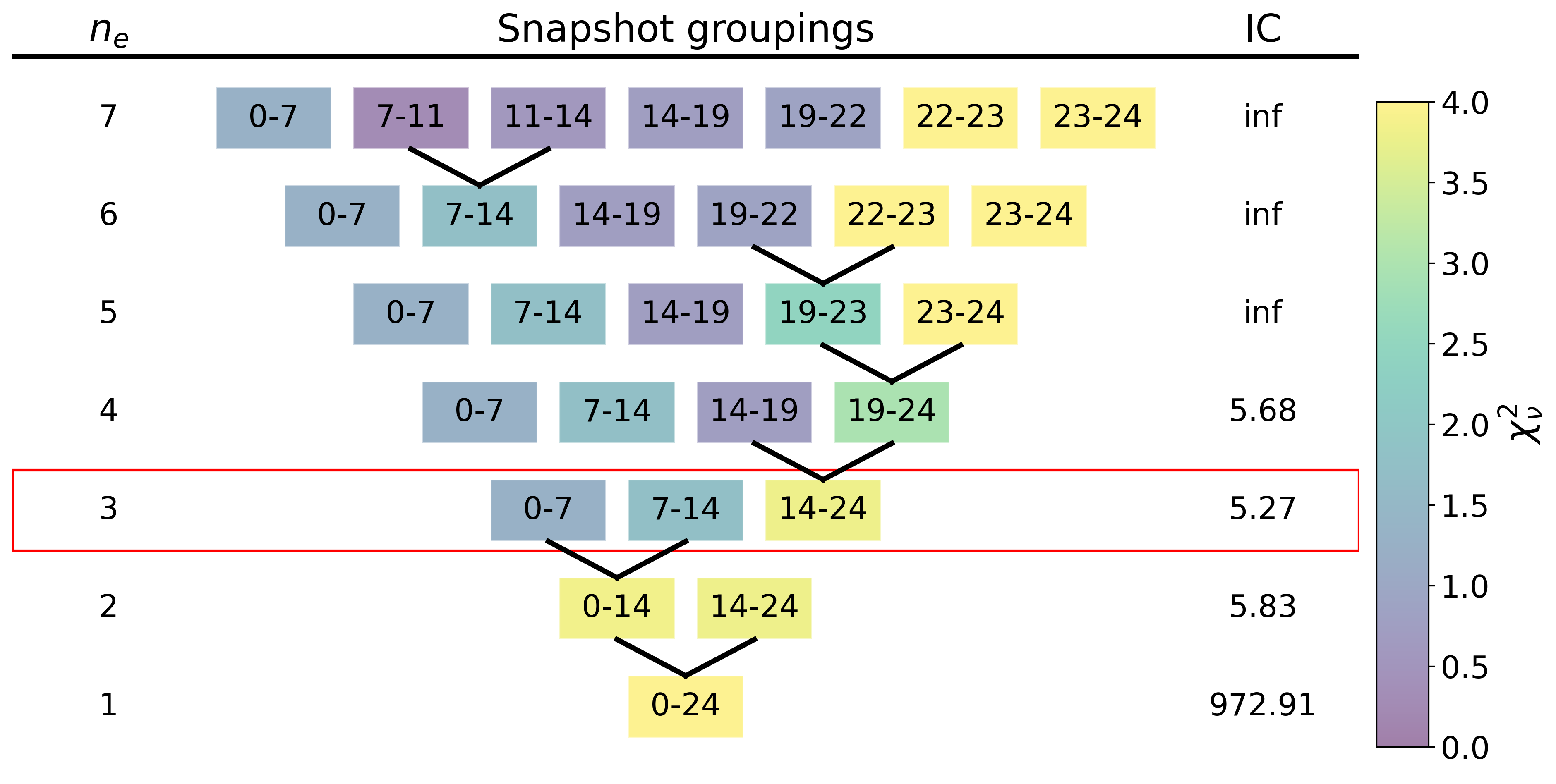}
    \includegraphics[width=0.365\textwidth]{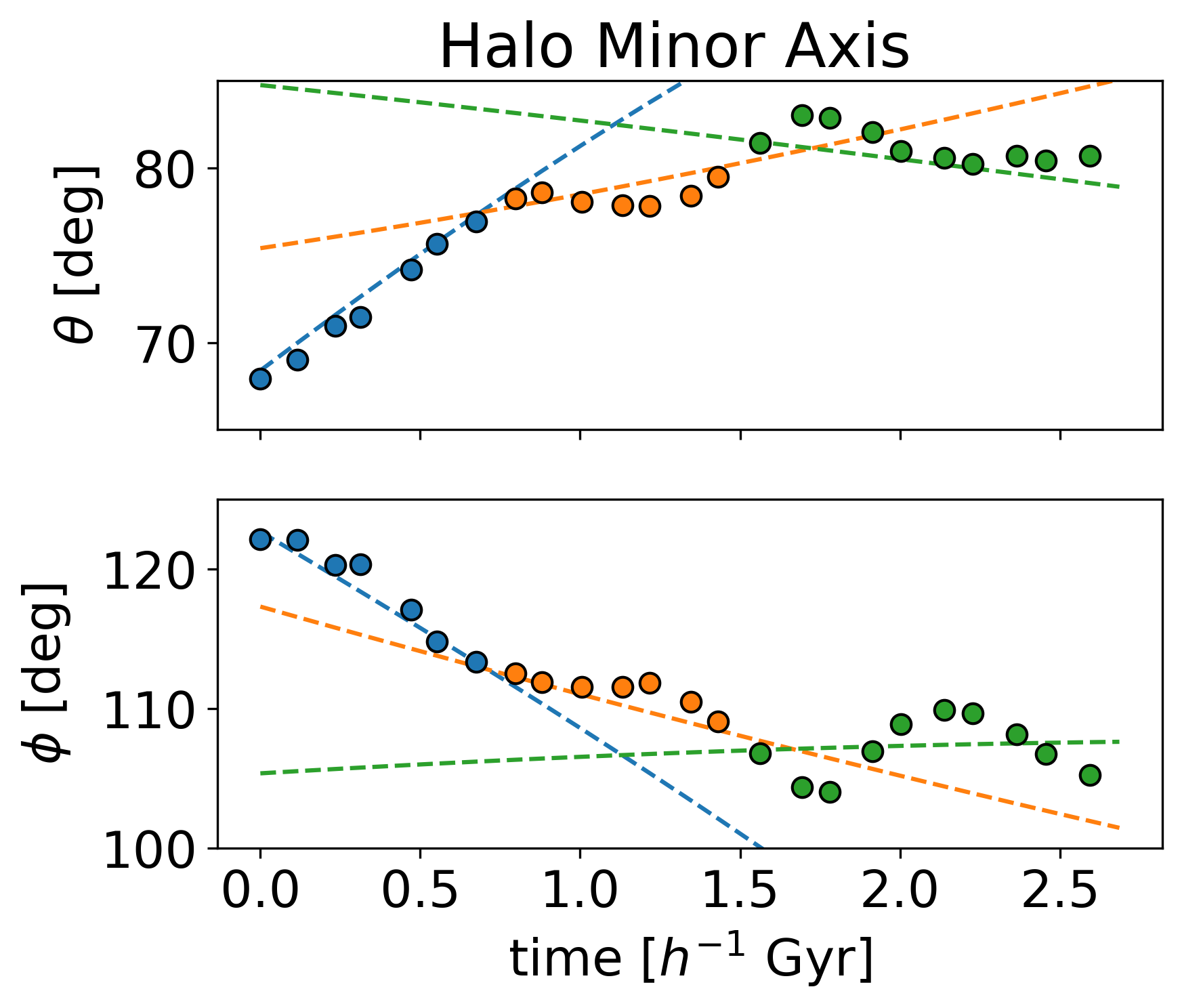}
    \caption{Example of the workflow outlined in section \ref{sec:rotation_epoch_determination}, showing the last 7 steps of the algorithm \added{for one sample halo (Left)}. At each step, a rotation axis fit is determined for each snapshot grouping (represented by snapshot numbers 0-24, corresponding to TNG50 snapshots 75-99), and a fit $\chi^2_\nu$ is measured using \ref{eq:single_axis_chi2}. The information criterion (IC) for all the snapshot groupings is measured using equation \ref{eq:IC}, and the next level is formed by merging the snapshot groupings which least increase the new fit $\chi^2_\nu$. The optimal snapshot groupings are selected as those with the minimum IC, which occurs in this example for $n_c=3$ implying that this halo has at most three coherent rotation epochs. The example halo shown here is the same halo used in figure \ref{fig:single axis fitting}. Note the reduced $\chi^2_\nu$ is undefined for snapshot groupings containing only 2 snapshots, because the degrees of freedom $\nu = 0$. \added{(Right) shows the evolution of the halo minor axis polar angle ($\theta$) and azimuthal angle ($\phi$) over time in the simulation box frame, colored by the rotation epochs found using our algorithm. The dashed lines show the angles predicted by the best fit planes and pattern speeds. The different epochs of rotation found by our algorithm are visible in these panels as changes in slope which disrupt the otherwise approximately linear evolution.}}
    \label{fig:genetic_algorithm}
\end{figure*}
\label{rotation epoch}

To determine the timescales over which given rotations remain stable, we propose a naive genetic algorithm. This algorithm compares different numbers of $n_e$ rotation epochs within a series of $n_s$ snapshots such that $1 \leq n_e < n_s $. For each rotation epoch, we measure the $\chi_\nu^2$ of the best fit rotation axis in order to assess an information criterion (IC) defined as:
\begin{equation}
    \text{IC} = n_e\Delta_{\chi^2} + \frac{1}{n_e}\sum_{i=1}^{n_e} \chi^2_i. 
    \label{eq:IC}
\end{equation}
The model achieving the minimum IC is selected, allowing for the optimal number of rotation epochs $n_e$ to be determined. This IC is defined such that a model with $n_e$ rotation epochs is preferred over a model with $n_e-1$ only if the reduced $\chi^2$ is lowered by an amount $\Delta_{\chi^2}$. We let $\Delta_{\chi^2}=1$. 

The objective of the algorithm is to determine both the optimal number of rotation epochs and their starting / ending snapshots. It proceeds as:
\begin{enumerate}
    \item Initialize $n_e = n_s - 1$ such that each pair of subsequent snapshots defines one rotation epoch. Measure the model IC.
    \item \label{gen_algo_2} Construct a model with $n_e = n_s - 2$ by merging the two neighboring rotation epochs which result in the smallest increase to the global fit $\chi^2$. Measure the model IC.
    \item Repeat step \ref{gen_algo_2} until $n_e = 1$.
    \item Select the model with the minimum IC.
\end{enumerate}
This algorithm allows us to approximate the optimal snapshot groupings for $n_e$ groupings without brute-force testing all arrangements. A brute-force computation would provide optimal snapshot grouping but is computationally intractable. Figure \ref{fig:genetic_algorithm} shows an example workflow of the algorithm from $n_e = 7$ to $n_e=1$. In practice, we run the algorithm for each halo over the full 25 snapshot time course.

\added{It is difficult and potentially impossible to exactly determine the boundaries between rotation epochs, in part because the simulation data is necessarily discretized into snapshots and because figure rotation is not a true solid-body rotation but rather is supported by certain orbits represented in the particle distribution, making the boundary between epochs fundamentally ambiguous. While we find our IC and algorithm provide a reasonable estimation of the boundaries between rotation epochs, other information criteria and algorithms may converge to answers which differ slightly from our own. For example, the left hand diagram of fig \ref{fig:genetic_algorithm} shows solutions with 4, 3, and 2 rotation epochs with very similar IC scores. While the minimum was found with 3 epochs, this suggests that 4 or 2 epochs could be reasonable solutions. However, a single epoch, or more than 4 epochs, can be ruled out by their large or divergent information criteria.}

\subsection{Pattern speeds}

The evolution of all three axes should be described by a single pattern speed $\Omega_p$ under an ideal rotation about a fixed, arbitrary axis. The angular coordinate $\phi_j$ of each principal axis $\hat{e}_j$ projected into its rotation plane will increase linearly in time according to: 
\begin{equation}
    \label{phi_v_t}
    \phi_j(t) = \Omega_p \cdot t + k,
\end{equation}
where $k$ is an arbitrary constant. The evolution of each $\phi_j$ may not follow a common pattern speed, in particular for cases where a principal axis is well aligned with the rotation axis. We therefore use an iterative method similar to the determination of rotation epochs to measure pattern speeds. The progression of $\phi_j$ for all three axes is first fit with a common pattern speed, according to equation~\ref{phi_v_t}. The axis contributing the most to the fit $\chi^2_\nu$ is cut, and a fit is performed with the remaining axis until only one axis remains. For $n_{axes} = 1,2,3$, we measure the information criterion
\begin{equation}
    \text{IC} = -10\cdot n_{axes} + \chi^2_\nu(n_{axes})
\end{equation}
and select the model $n_{axes}$ which minimizes the information criterion. This is condition is equivalent to dropping an axis from the fit only when it improves $\chi_\nu^2$ by 10 or more. We consider the fit to be a non-detection of figure rotation if  $\chi_\nu^2 < 20$ is not achieved with $n_{axis} = 1$. We repeat this fitting procedure for all halos at all snapshot groupings defined using the methods of section \ref{rotation epoch}.

\section{Results}\label{sec:results}

\begin{figure}
    \centering
    \includegraphics[width=0.45\textwidth]{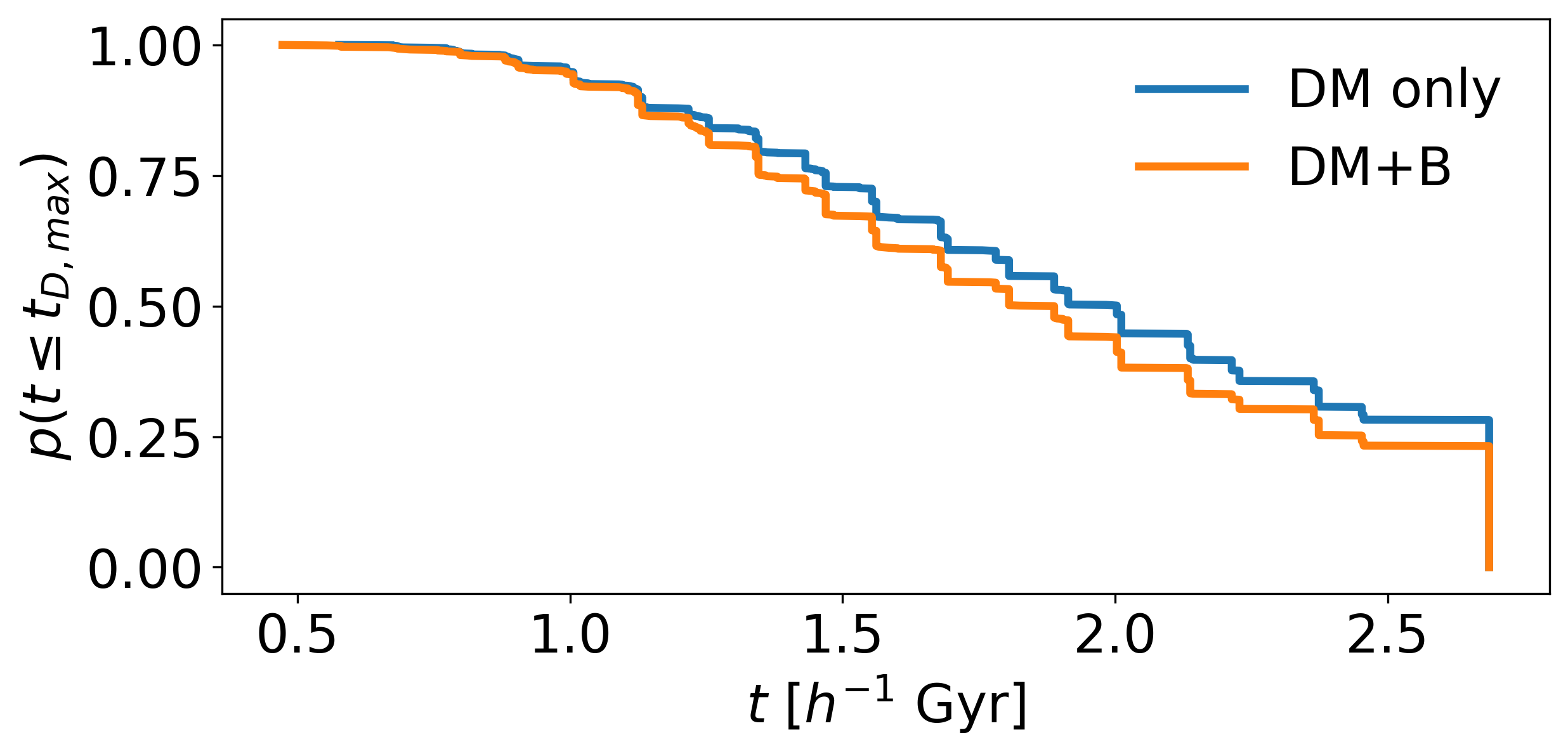}
    \includegraphics[width=0.45\textwidth]{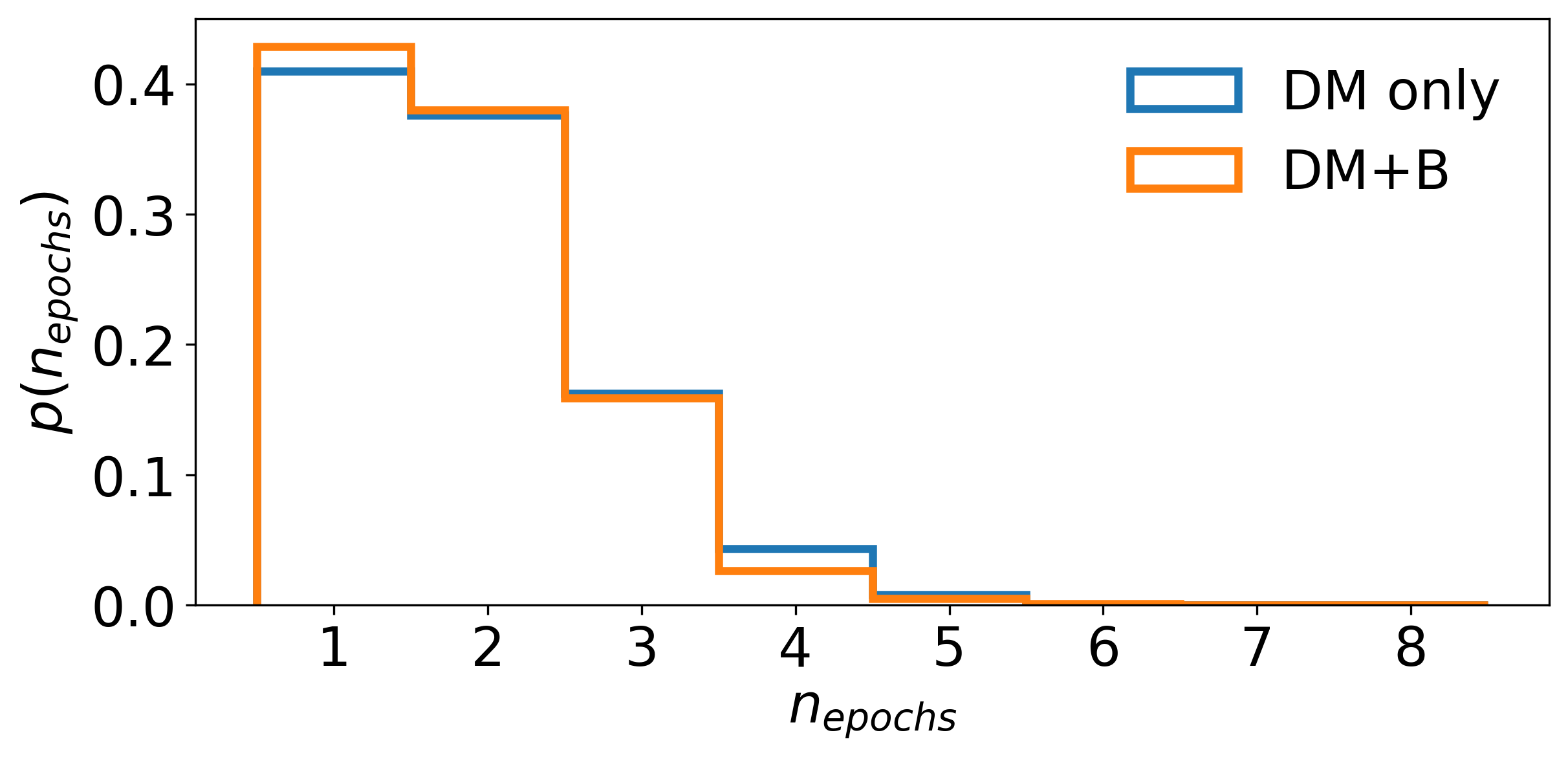}
    \caption{\added{(Top) Reverse-ordered cumulative histogram showing the fraction of halos whose longest-lived rotation is at least as large as a given duration. This fraction drops faster for halos in the DM+B run of TNG50, indicating a slightly reduced abundance of long-lived rotations compared to the DM only case. (Bottom) histogram of the number of rotation epochs found for each TNG50 run. The number of found epochs is similar for both runs, suggesting that the differences in the top panel histograms cannot be explained by a difference in the number of rotation epochs.} }
    \label{fig:duration_hist}
\end{figure}

\begin{figure*}
\centering
\includegraphics[trim={20 60 0 0},clip, width=0.4\textwidth]{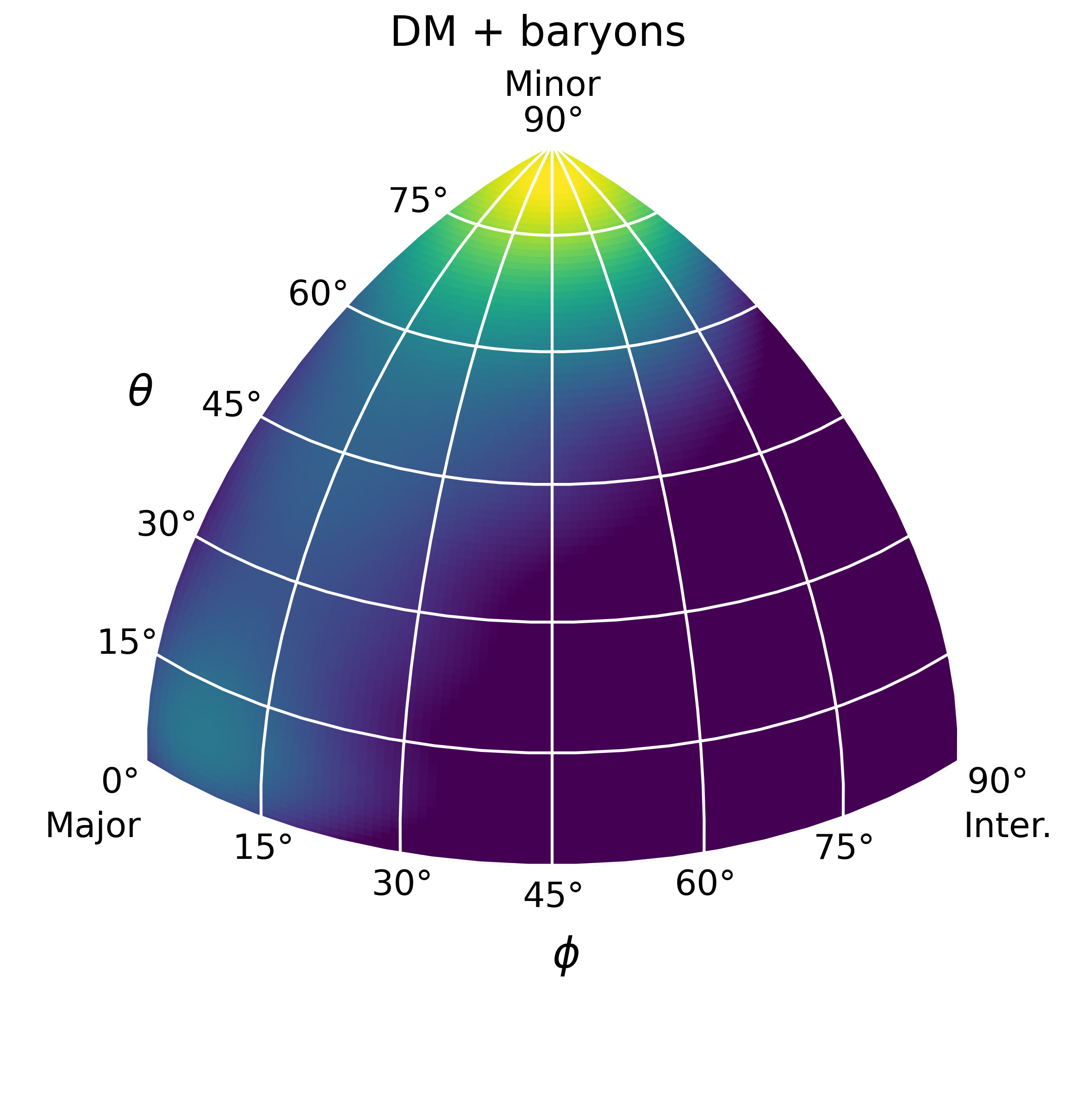}
\includegraphics[trim={20 60 0 0},clip, width=0.48\textwidth]{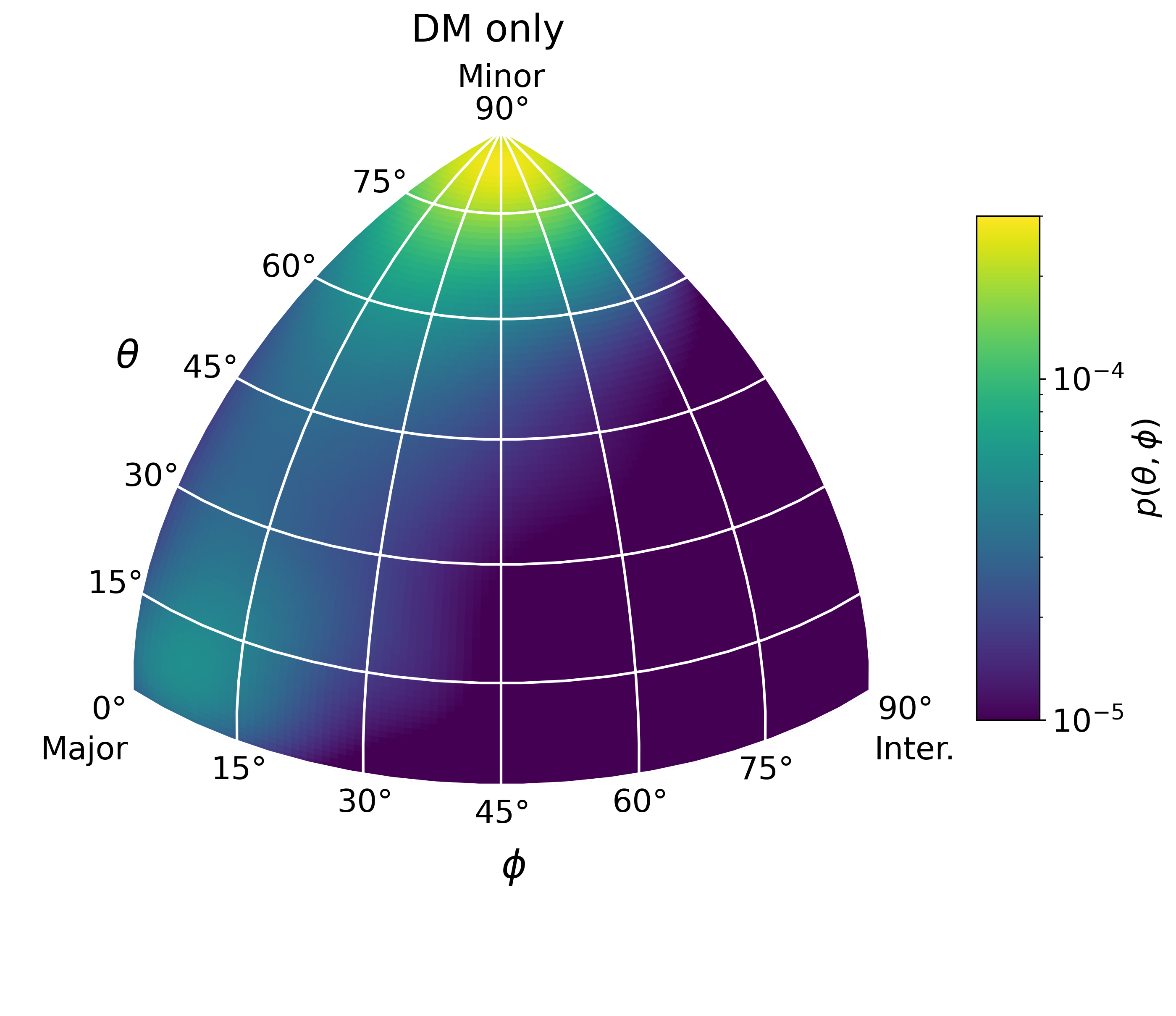}
\includegraphics[trim={20 80 0 20},clip, width=0.4\textwidth]{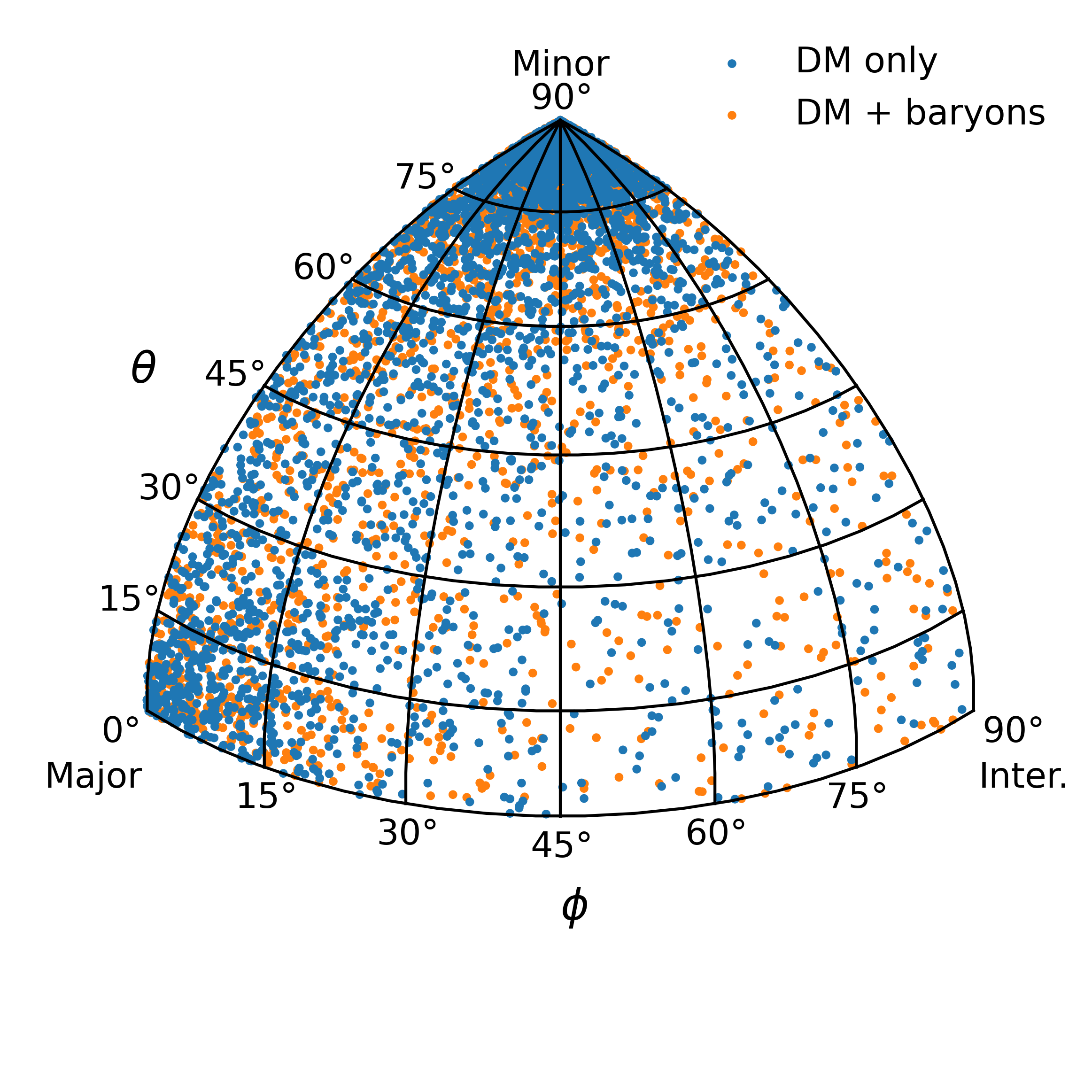}
\caption{Normalized 2D histograms of rotation axis orientations for mixed DM+B runs (top left) and DM-only (top right) TNG50 runs. Bottom row shows scatter plot of the same halos. The halo minor axis is along the $z$ direction ($\theta = 90^\circ$) and the major axis is along the $x$ direction ($\theta, \phi = 0^\circ$). DM-only and DM+B simulations show qualitatively similar results, with a majority halos having  their rotation axes aligned with the halo minor axis or major axis. Rotation axes which are aligned with no principal axis are present in both types of simulations.}
\label{fig:octant_prob}
\end{figure*}

\begin{figure*}
    \centering
  \includegraphics[width=0.44\textwidth]{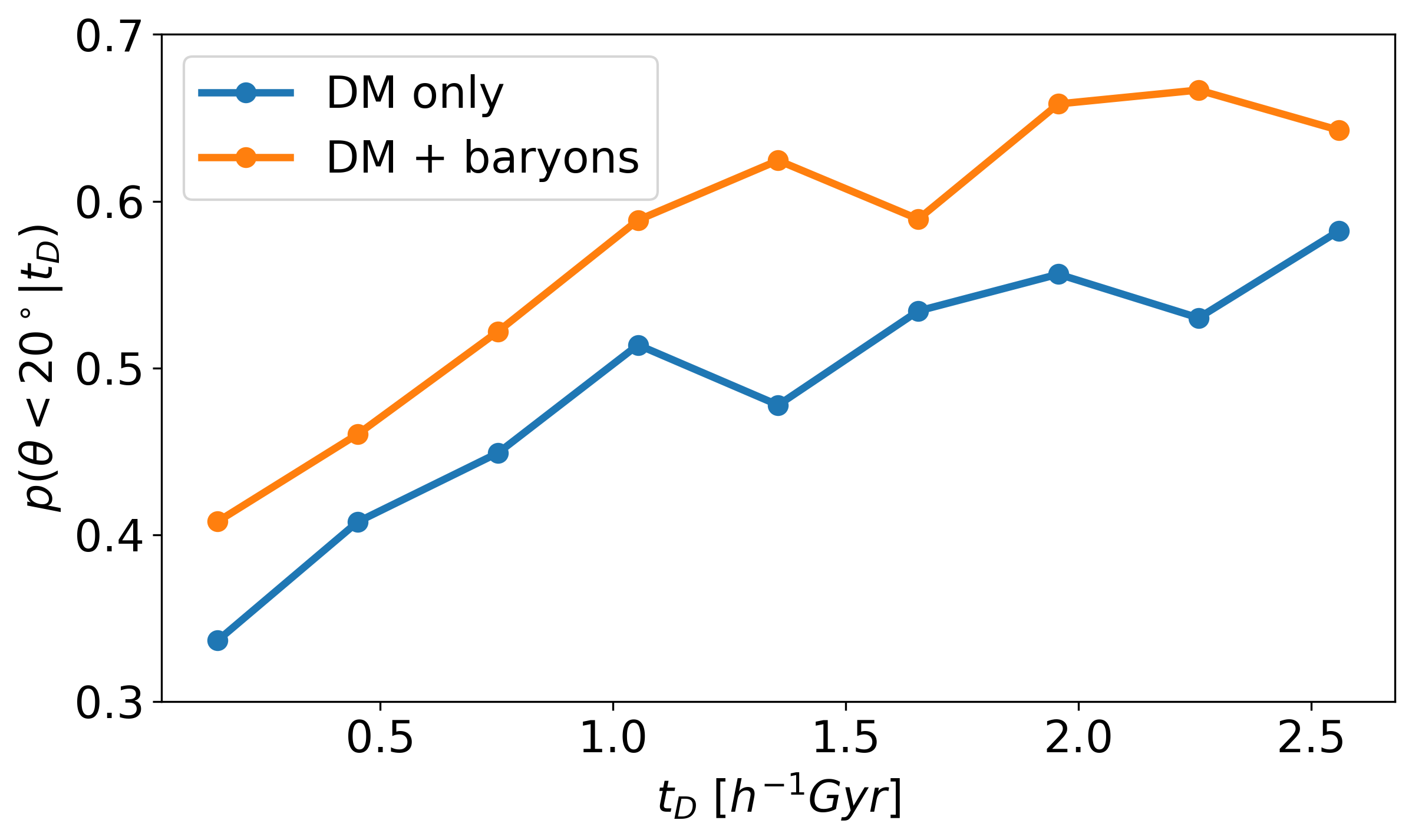}
    \includegraphics[width=0.46\textwidth]{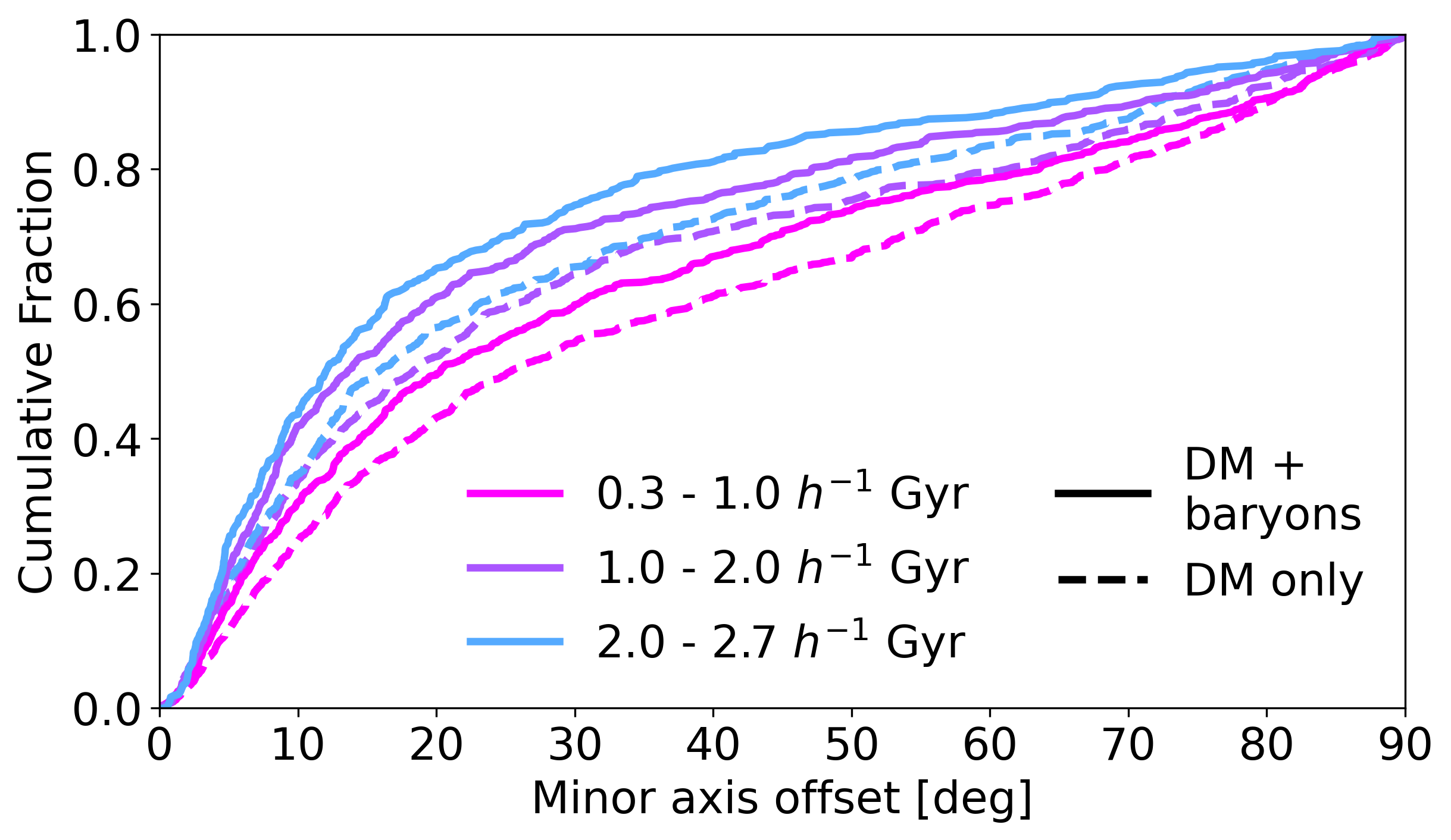}
  
\caption{\textbf{Left:} Fraction of figure rotation axes within $20^\circ$ of the halo minor axis against the duration of the rotation. \textbf{Right:} Cumulative distribution of the rotation axis polar angle, binned in rotation duration. A higher cumulative fraction of rotation axes within a given angular distance indicates stronger alignment between the figure rotation axis and the halo minor axis. In both DM-only and DM+B TNG50 realizations, longer-lived rotations show a higher degree of alignment between the rotation axis and the halo minor axis. In the presence of baryons, rotation axes are better aligned with the halo minor axis at all rotation durations.  }
  \label{fig:minor_axis_alignment_v_tD}
\end{figure*}

In the following section, we outline the results we obtain using the rotation axis fitting method described in section \ref{sec:rotation axis fitting}. We consider a detection of figure rotation to be positive when:
\begin{enumerate}
    \item\label{criterion_axis_chi2} The rotation axis fit (eq. \ref{eq:single_axis_chi2}) $\chi^2_\nu < 10$. $\chi^2_\nu$ exceeding $10$ for a given epoch indicates motion in the halo which cannot be described as a simple rotation.
    \item\label{criterion_phi_v_t_chi2} The $\phi_j(t)$ fit $\chi_\nu^2 < 20$. $\chi^2_\nu$ for the $\phi_j(t)$ were occasionally large due to small ``nodding'' movements in the axes, possibly arising from precession. We allow the more relaxed $\chi^2_\nu$ requirement here so as to include these halos in our final selection.
    \item\label{criterion_rotation_above_lower_bound} The angle swept out by a principal axis during a rotation is greater than its angular uncertainty.
    \item\label{criterion_duration} The duration of the rotation is $> 3$ snapshots. \added{This sets a minimum detectable duration of $\sim330 h^{-1}$ Myr.}
    \item\label{criterion_2_sigma} The pattern speed is detected above the $2\sigma$ threshold. 
\end{enumerate}
Criterion \ref{criterion_rotation_above_lower_bound} and \ref{criterion_duration} are enforced to eliminate spurious fits. Criterion~\ref{criterion_rotation_above_lower_bound} in particular is enforced to eliminate cases in which random-walk motions due to the imperfect measurement of the principal axis orientations creates a false signal. This is distinct from criterion \ref{criterion_2_sigma}, which is achieved when the measured pattern speed is at least twice the uncertainty on the fit pattern speed, as measured by the fit covariance. We enforce \ref{criterion_2_sigma} to ensure that all halos allowed by the relaxed $\phi_j(t)$ fit $\chi_\nu^2$ of requirement \ref{criterion_phi_v_t_chi2} represent true figure rotation.

\subsection{Prevalence of figure rotation}

We find a positive detection of figure rotation in 1,484 of the 1,577 DM-only halos ($\sim 94.1\%$) and 1,139 of the 1,396 ($\sim 81.6\%$) of the DM + baryon halos within the sampled time course. For the set of halos with positive detections of figure rotation, $94.8\%$ of DM-only halos and $94.4\%$ of DM+B halos had figure rotation lasting for at least $1~ h^{-1} $ Gyr. For durations lasting longer than $2~ h^{-1} $ Gyr, the fractions drop to $50.1\%$ for DM-only halos and $44.1\%$ for DM+B halos. We additionally find 419 ($28.2\%$) DM-only halos and 265 ($23.3\%$) DM+B halos which undergo coherent figure rotation for the full simulation duration. The duration of these rotations is hence $\geq 2.7~h^{-1}$ Gyr, but cannot be precisely determined. In the population of halos with positive detections of figure rotation, the median number of steady rotation epochs was 2 for both simulations. \added{The bottom panel of figure \ref{fig:duration_hist} further shows that the distribution of the number of rotation epochs was nearly identical between the DM+B and DM only runs.}

Taken together these results show that figure rotation over periods of $1~ h^{-1} $ Gyr in our halo catalogs are common in both DM-only and DM+B halos, but slightly less common in DM+B halos relative to their DM-only counterparts. Notably, the fraction of the 1,577 DM-only halos within our surveyed sample which undergo figure rotation for at least $1~ h^{-1} $ Gyr was $ 89.2\%$, in good agreement with \cite{bailin_figure_2004} but higher than the $61\%$ found by \cite{bryan_figure_2007}. The fraction for the 1,396 DM+B halos is $77.0\%$, showing that figure rotation over these durations is common in the presence of baryons, but slightly reduced relative to the figure rotation in the absence of baryons. \added{The top panel of figure \ref{fig:duration_hist} shows the observed fraction of rotations with durations above a given value. For durations longer than $\sim 1.2 h^{-1}$ Gyr, the fraction of rotations sustained for those durations or longer is reduced in the DM+B run relative to the DM only run. These results suggest that both the overall fraction of rotating halos is lower and that longer-lived rotations are less common in the DM+B simulation. The bottom panel of \ref{fig:duration_hist} further suggests that the difference in durations between runs cannot be explained by a greater number of short-lived rotations in the DM+B simulation.}

\subsection{Orientation of principal axes}

We next explore the 2-dimensional distributions of the figure rotation axes. Figure \ref{fig:octant_prob} shows 2D histograms and a scatter plot of the observed rotation axis orientations, relative to the three principal axes. Qualitatively, our results are similar between the two simulations. In each case, the figure rotation axes are predominantly aligned with the halo minor axis ($47.9\%$ are aligned within $20^\circ$ for the DM-only  simulation and $54.7\%$ for the DM+B simulation) or with the major axis ($8.7\%$ for DM only and $6.9\%$ for DM+B). The alignment with the major axis is somewhat weaker and the alignment with the minor axis is somewhat stronger in DM+B halos as compared to the DM-only. 

We observe very few halos in either type of simulation with rotation axes aligned near the halo intermediate axis. This result is probably expected: tube orbits and planar orbits are unstable when their angular momentum is aligned with the intermediate axis of a triaxial potential \citep{heiligman_nonexistence_1979,goodman_semistochastic_1981,wilkinson_stationary_1982,adams_orbital_2008,carpintero_three_2012}. Simulations have shown that disk galaxies initialized to rotate about the intermediate axis will flip over to reorient themselves closer to the minor or major axis of the halos \citep{debattista_whats_2013}. Each of these gives some reason to expect that figure rotation about the intermediate axis could be a less stable configuration. Such an instability would be reminiscent of the classical rotational instability of solid bodies about their intermediate axis \citep[e.g.][]{taylor_classical_2005}. In our catalogs, we find a small but nonzero fraction of halos with figure rotation axes aligned within $20^\circ$ of the halo intermediate axis. This fraction was $0.7\%$ and $0.6\%$ for DM-only and DM+B halos, respectively. These results are consistent with both \cite{bailin_figure_2004} and \cite{bryan_figure_2007}, who observed no halos with figure rotation about their intermediate axis but studied $\sim 300$ halos. Based on our observed abundance, we would not expect to observe rotation about the intermediate axis in samples smaller than $\sim$ 1000 halos.

Intriguingly, the fractions of figure rotation axes aligned with the halo intermediate axis does not drop to zero for the longest-lived rotations in either the DM-only simulation or the DM+B simulation as one would expect if they were completely unstable. In our catalogs we observe 3 DM-only halos and 1 DM+B halo with rotation axes $\leq 20^\circ$ from the halo intermediate axis whose rotations were stable over the full $2.7 ~h^{-1}$ Gyr period surveyed. This may suggest that, while very rare, stable figure rotation with a rotation axis near the intermediate axis is possible.

Rotation axes which are not aligned with either the minor or major halo axes are predominantly located within $\sim 30^\circ$ of the plane containing the minor and major axes, as seen in figure \ref{fig:octant_prob}. We find that such halos can make up a substantial fraction of the full population. Over all durations, we find that $42.7\%$ of DM-only halos and $37.8\% $ of DM+B halos have rotation axes not aligned within $20^\circ$ of any principal axes. These rotations were observed to be stable over the surveyed time course. Among the halos with rotation lasting $\geq 2.7~h^{-1}$ Gyr, we find that the fraction of rotation axes misaligned with any principal axis are $35.1\%$ for DM-only halos and $33.2\%$ for DM+B halos. These fractions are relatively large and suggest that halos with a figure rotation axis not aligned with any halo principal axis are not uncommon and could have important dynamical implications (e.g. they  could induce warps in disks and other baryonic components such as tidal streams).

Figure \ref{fig:minor_axis_alignment_v_tD} shows the evolution of alignment with the minor axes with respect to the duration of rotation. We present this in two alternate ways; the left panel of figure \ref{fig:minor_axis_alignment_v_tD} shows the fraction of figure rotation axes within $20^\circ$ of the halo minor axis against the duration of the rotation. The right panel of figure \ref{fig:minor_axis_alignment_v_tD} shows the cumulative fraction of rotation axes within a given angular distance of the halo minor axis, for rotations binned in durations of $0.3~h^{-1}-1~h^{-1}$ Gyr, $1~h^{-1}-2~h^{-1}$ Gyr, and $2~h^{-1} - 2.7~h^{-1}$ Gyr. Each of these panels demonstrates that longer lived rotations tend to show a higher degree of alignment with the halo minor axis, though rotation axes which are not aligned with the halo minor axis still persist for the longest durations $\gtrsim 2.7~h^{-1}$ Gyr. Alignment with the halo minor axis is stronger in the DM+B halos relative to the DM only population across all durations. Interestingly, the longest-lived rotations show a systematic trend towards better alignment with the halo minor axis, both in the presence and absence of baryons. We interpret this as evidence that rotations aligned with the halo minor axis are more stable than those which are aligned with the halo major axis or are not aligned with any principal axis. Because the offset between alignments in the presence and absence of baryons is persistent across all durations, we can rule out the difference in mean durations between the DM-only and DM+B simulations as an explanation for the greater alignment with the minor axis in the presence of baryons. 

The minor axis alignment we observe is lower compared to that found by \cite{bailin_figure_2004}, who report the alignment to be $\sim 85\%$, but is comparable to the results of \cite{bryan_figure_2007} who find the value to be closer to $50\%$.

\begin{figure*}
    \includegraphics[width=.575\textwidth]{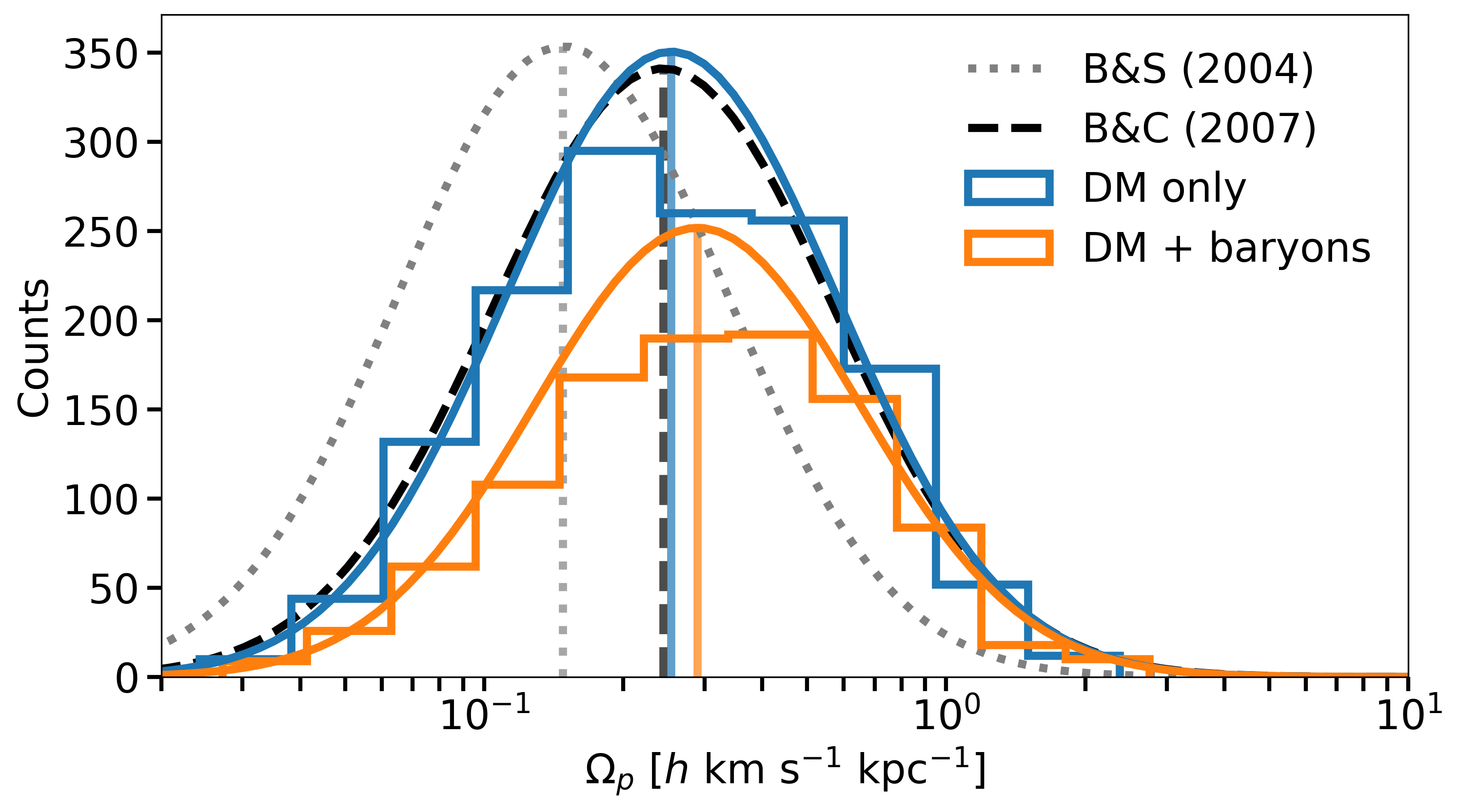}
    \includegraphics[width=.425\textwidth]{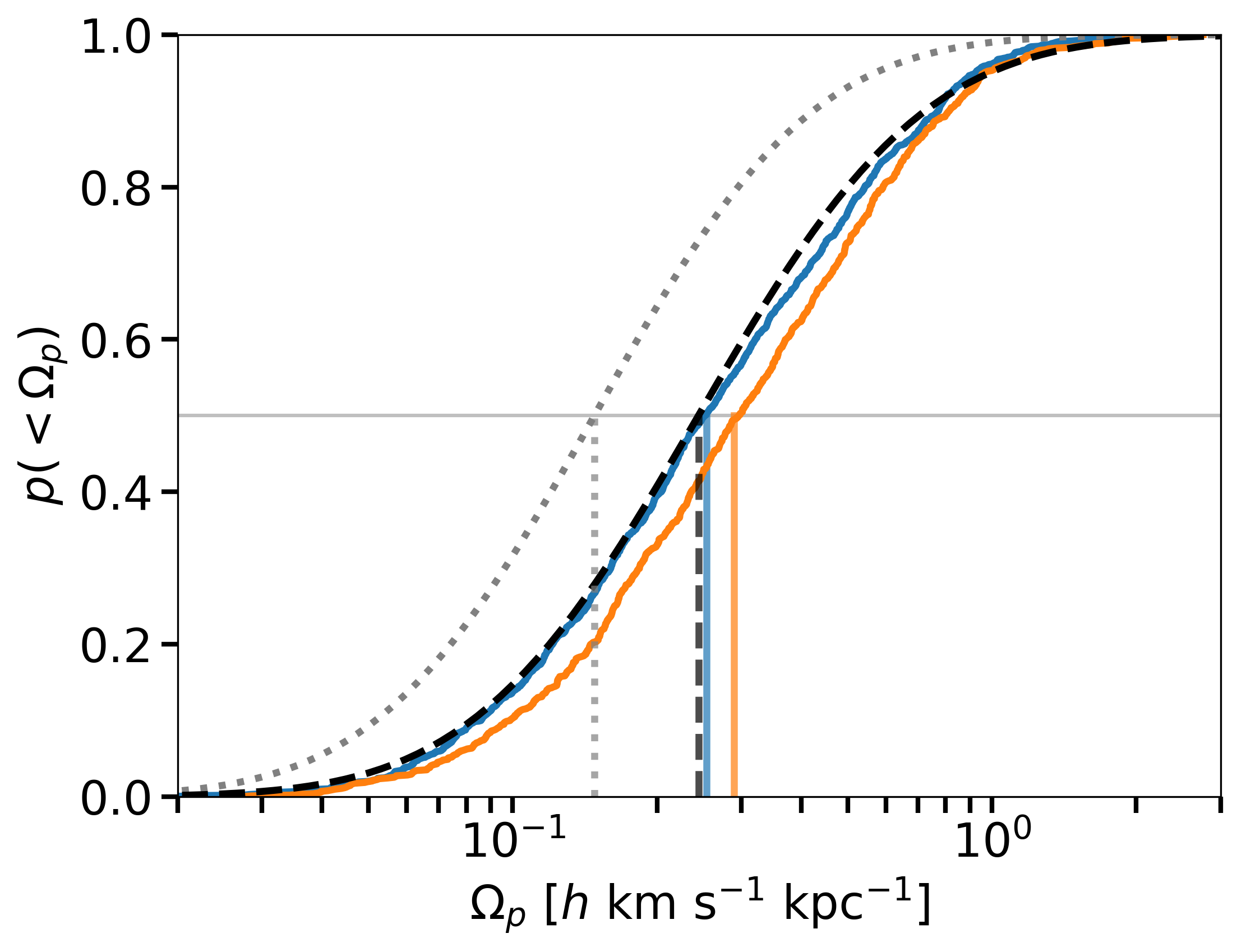}
    \caption{Measured pattern speeds for rotations lasting $\geq$ 1 $h^{-1}$Gyr for DM-only and DM+B runs of TNG50. The histograms are fit by log-normal distributions with a mean of 0.254 $\pm$ 0.005 $h$ km s$^{-1}$ kpc$^{-1}$ for the DM-only simulation and 0.289 $\pm$ 0.007 $h$ km s$^{-1}$ kpc$^{-1}$ for the mixed-phase DM+baryonic run. Predicted counts for our DM-only halos using the distributions measured by \cite{bailin_figure_2004} and \cite{bryan_figure_2007} are shown. There is very good agreement between our results and those of \cite{bryan_figure_2007} in TNG50 DM-only halos, but observe a slight shift to higher pattern speeds in the TNG50 DM+B halos of order $\sim 14\%$.}
    \label{fig:pattern_speed_hist}
\end{figure*}

\subsection{Pattern speed distributions}\label{sec:pattern_speed_dist}

In the sampled population of halos, there are 1,454 distinct epochs of figure rotation lasting longer than $1 h^{-1}$ Gyr in the DM-only halos, and 1023 in the DM+B halos. Our interest in this work is the  characterization of figure rotation during epochs of coherent rotation. Because individual halos may undergo many distinct epochs of coherent rotation, halos may appear in this catalog at most twice since the full time course covers $2.7h^{-1}$ Gyr. In a future work we will investigate what causes these distinct epochs of figure rotation. The distribution of pattern speeds we measure in these halos is shown in figure \ref{fig:pattern_speed_hist}. Our pattern speeds are well-described by a lognormal distribution for halos in both the DM-only and DM+B simulations:
\begin{equation}
    p(\Omega_p) = \frac{1}{\Omega_p\sqrt{2\pi\sigma^2}} \text{exp}\left( -\frac{\text{ln}(\Omega_p/\mu)^2}{2\sigma^2} \right),
    \label{eq:logNormal}
\end{equation}
where $\sigma$ gives the natural width and $\mu$ gives the median of the distribution. We estimate the optimal values of $\sigma$ and $\mu$ by running an expectation-maximization fitting routine on 500 bootstrap samples, which we repeat independently for both the DM-only and DM+B simulations. The distributions we obtain from this procedure are seen overplotted on figure \ref{fig:pattern_speed_hist}, and are compiled in table \ref{tab:PS_fit_params}. At nearly $5\sigma$ confidence, the best-fit median pattern speed for the DM+B halos was found to be $14\%$ higher at $0.29 h$ km s$^{-1}$ kpc$^{-1}$ compared to the DM-only halos at $0.25 h$ km s$^{-1}$ kpc$^{-1}$. The observed median pattern speeds for both simulations were higher than those found by \cite{bailin_figure_2004} by  $\sim 72\%$ and $95\%$, but the DM-only median closely matches the result of \cite{bryan_figure_2007} to within $4\%$. The widths of our distributions are $0.83 h$ and $0.81 h$ km s$^{-1}$ kpc$^{-1}$ for the DM-only and DM+B runs, in close agreement with the widths found by both \cite{bailin_figure_2004} and \cite{bryan_figure_2007}. Our results for both the median and width for the DM-only case match with those of \cite{bryan_figure_2007} to better than $1\sigma$.
\begin{table}
\begin{center}
\begin{tabular}{c| c c} 
   & $\mu$  & $\sigma $  \\ 
 \hline
 DM-only & $0.254 \pm 0.005$ & $0.83 \pm 0.01$ \\
 DM + baryons & $0.289 \pm 0.007$ & $0.81 \pm 0.02$ \\
 \cite{bryan_figure_2007} & $0.24 \pm 0.02 $ & $0.85 \pm 0.07$ \\
 \cite{bailin_figure_2004} & $0.148$ & $0.83$  \\
\end{tabular}
\caption{Best-fit parameters of the lognormal distribution for pattern speeds lasting $\geq 1 h^{-1}$ Gyr, as in equation \ref{eq:logNormal}. Comparisons to \cite{bryan_figure_2007} and \cite{bailin_figure_2004} are shown. All values are reported in $h$ km s$^{-1}$ kpc$^{-1}$.}
\label{tab:PS_fit_params}
\end{center}
\end{table}

To assess the significance of the difference between the DM-only and DM+B pattern speed distributions, we perform an independent two-sample $t$-test on the base 10 logarithms of our measured pattern speeds. The logarithm of our pattern speeds are normally distributed with approximately the same distribution width. The $t$-test is written as 
\begin{equation}
    \frac{1}{s_P}\sqrt{\frac{n}{2}}(\bar{X}_1 - \bar{X}_2), \quad s_P = \frac{1}{2}\sqrt{s_{X_1}^2 + s_{X_1}^2}
    \label{t-test}
\end{equation}
Where $\bar{X}$ represents the observed population mean and $s_X$ is the observed population standard deviation. Using these definitions, we calculate a $t$ statistic of $-3.93$ between the DM-only and DM+B populations, showing the higher mean of the DM+B population. This $t$ statistic returns an achieved significance level ($p$-value) of  $8.76\times10^{-5}$. This suggests that it is very unlikely the pattern speeds in our DM-only and DM+B halos can be described by a single log-normal distribution, and suggests that the $14\%$ higher median pattern speed observed in the DM+B halos is robust. 

We find that the pattern speeds in our halo catalog show no dependence on halo mass throughout the mass range probed by our catalog of $\sim 10^{10} - 10^{12} h^{-1}$ M$_\odot$ (see Appendix \ref{appendix:mass_pattern_speed}). This result is in agreement with \cite{bailin_figure_2004}, and probes a much lower mass range than was previously available. We therefore conclude that the different mass ranges covered by our study and that of \cite{bailin_figure_2004} can not explain the offset of in our observed pattern speed distributions.

\begin{figure*}
    \centering
    \includegraphics[width=.9\textwidth]{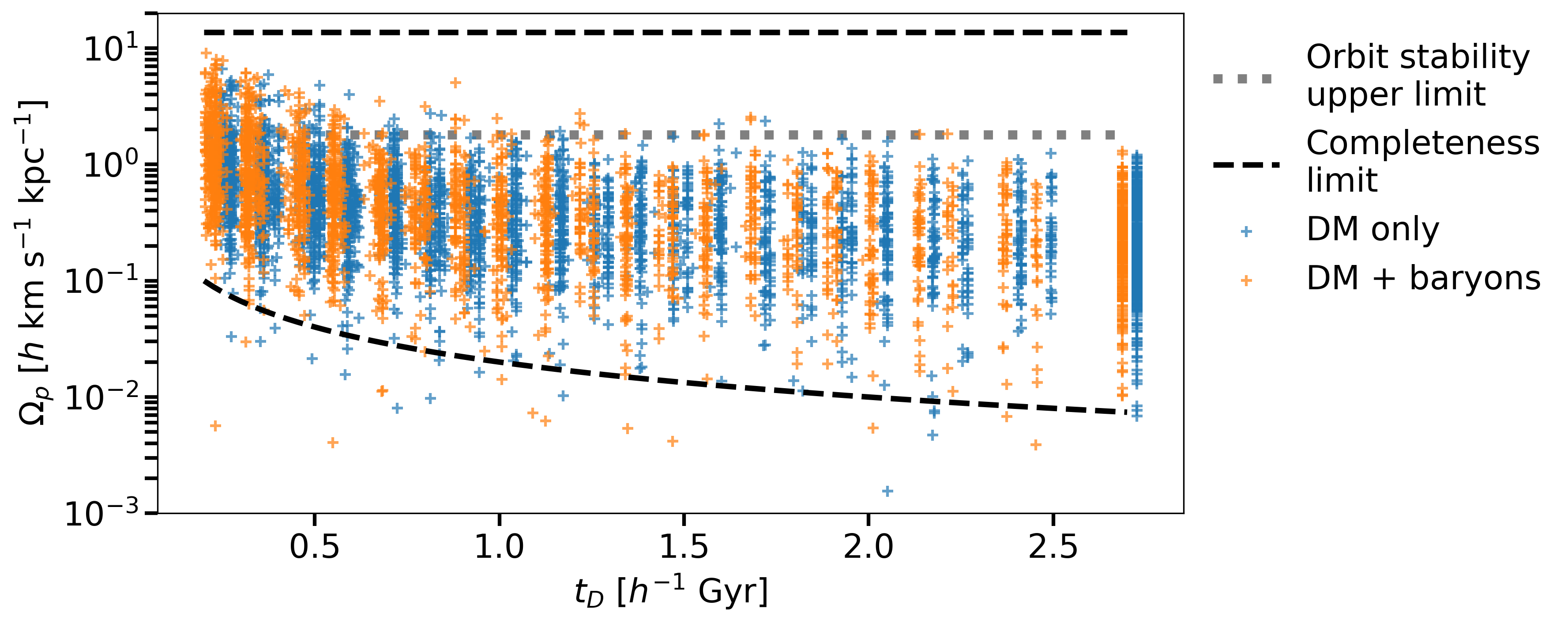}
    \caption{Figure rotation pattern speeds against rotation durations. For visibility, the durations of DM-only halos  have been offset by +0.04 $h^{-1}$ Gyr. Pattern speeds decrease with longer-lived rotations out to durations of $\sim 1~h^{-1}$ Gyr, beyond which they stabilize. The upper bound of the distribution past $1~h^{-1}$ Gyr closely matches the upper stability limit of figure rotation found by \cite{deibel_orbital_2011}. Completion limits are shown as black dashed lines. The lower completion bound is set by the pattern speed necessary for the principal axes to sweep out an angle greater than the typical uncertainty on the axis orientation, for a given duration. Pattern speeds below the lower completion limit are considered non-detections and are omitted. The upper bound is set by the Nyquist frequency of $90^\circ$/snapshot. Short-lived rotations ($t_D \lesssim 1 h^{-1}$Gyr) can have higher pattern speeds relative to long-lived rotations.}
    \label{fig:pattern_speeds_vs_duration}
\end{figure*}

\subsection{Pattern speeds vs. duration of figure rotation}

In figure \ref{fig:pattern_speeds_vs_duration} we show the behavior of pattern speeds with respect to their durations. We observe a systemic trend to lower pattern speeds with increasing duration of rotation in both the presence and absence of baryons up to durations of $\sim 1 h^{-1}$ Gyr, at which point the pattern speeds begin to stabilize. At the lowest durations, the pattern speed distribution approaches the completeness limits on both the upper and lower ends. The lower edge of the pattern speed distribution remains near the lower completeness limit until $\sim 1 h^{-1}$ Gyr. We therefore have not constrained the lower bound of the pattern speed distribution for short durations, however we note that beyond the very shortest durations the upper edge of the distribution is well separated from the upper completeness limit. Here, the upper completeness limit is the Nyquist frequency of $90^\circ$/ snapshot ($\sim 10 h - 14 h$ km s$^{-1}$ kpc$^{-1}$). The lower completeness limit is given by the angular uncertainty on the principal axis orientations, which places a lower bound on the detectable pattern speeds as outlined by criterion \ref{criterion_rotation_above_lower_bound} specified at the beginning of section \ref{sec:results}.

We emphasize that this plot does not show pattern speeds slowing down over the lifetime of their rotation, rather it is suggestive that slower rotations tend to be longer-lived. This trend may hint at an instability present for fast-rotating halos. This picture is consistent with that presented by \cite{deibel_orbital_2011}, who find that rotational periods shorter than $\sim 5$ Gyr (with corresponding pattern speeds above $\sim 1.8 h$ km s$^{-1}$ kpc$^{-1}$) destabilize the orbits necessary to maintain a steady triaxial shape. This pattern speed is close to the upper limit we observe for rotations with lifetimes greater than $\sim 1 h^{-1}$ Gyr. 

\subsection{Stability of figure rotation}

\begin{figure*}
    \centering
      \includegraphics[trim={20 60 110 0},clip, width=0.4\textwidth]{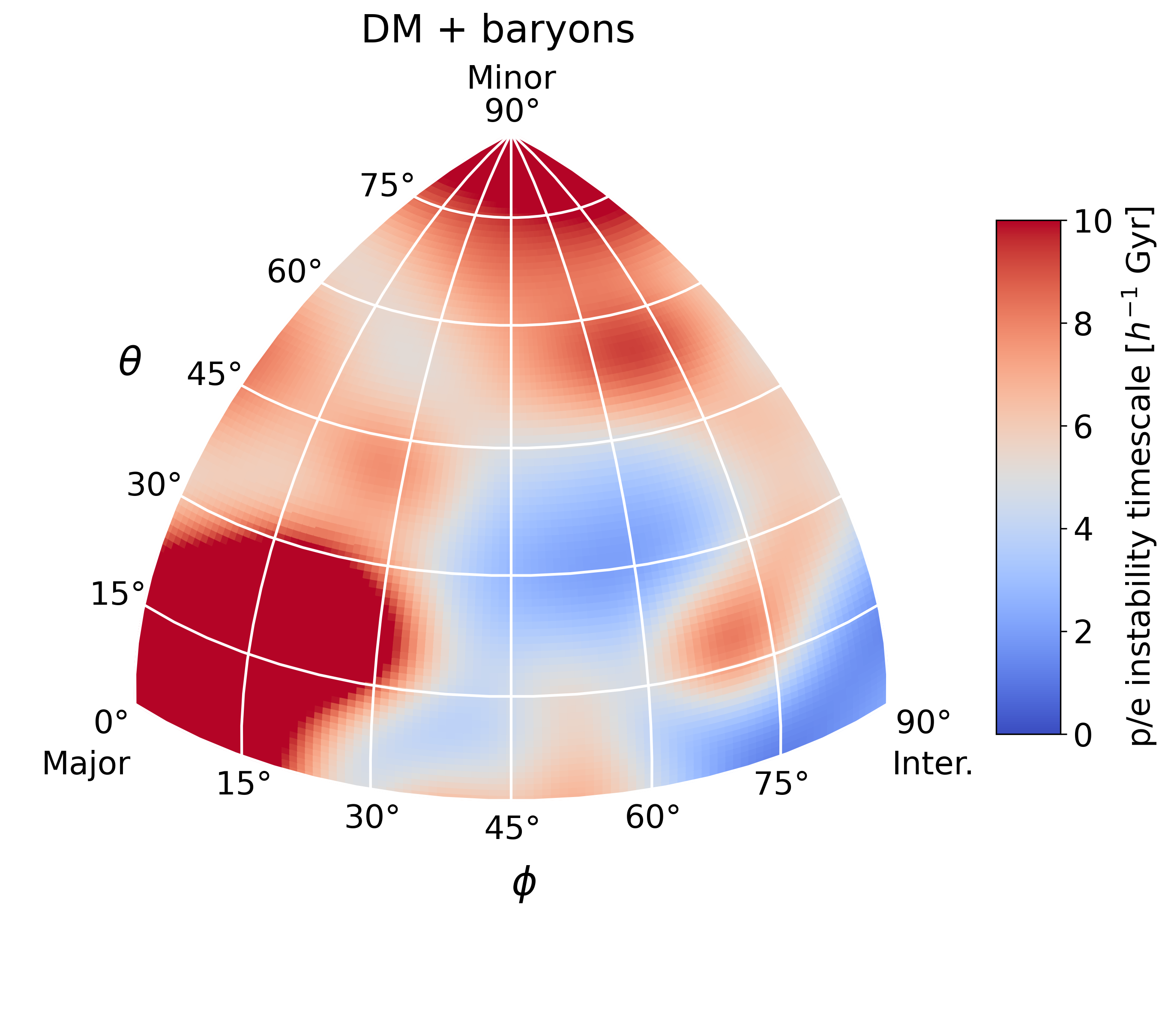}
      \includegraphics[trim={20 60 0 0},clip, width=0.5\textwidth]{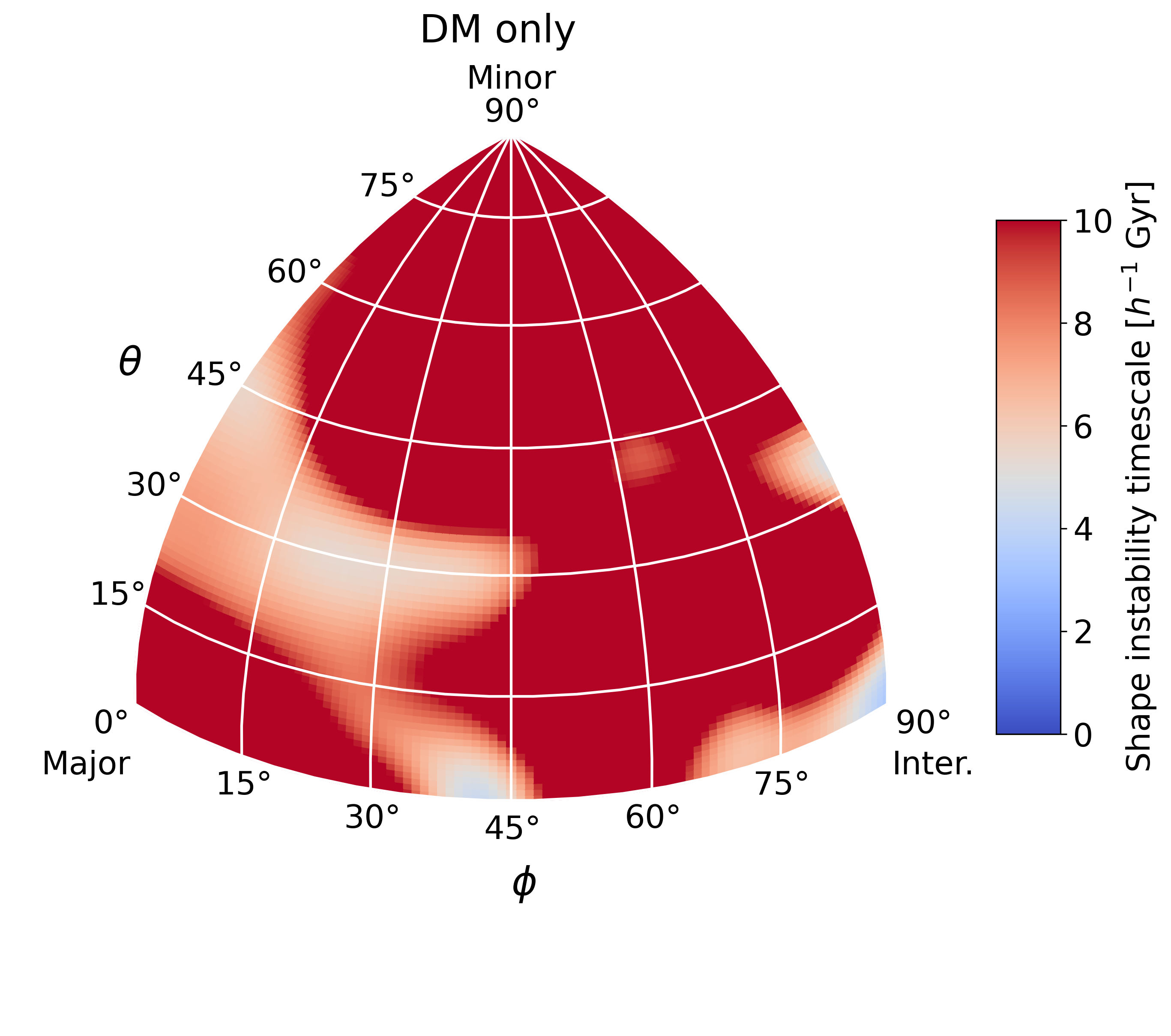}
    \caption{2D histograms showing locally averaged timescales for significant evolution of the halo shape as a function of the rotation axis orientation. Timescales are calculated as the halo $p/e$ shape divided by the average time derivative of $p/e$ during the period of rotation. Local averages are calculated using a Gaussian kernel with $\sigma = 6^\circ$. In both DM only and DM + baryon simulations, halos undergoing figure rotation about the minor or major axis have shapes which are stable for greater than one Hubble time, while halos undergoing figure rotation about the intermediate axis have shapes which may change significantly within $\sim 2 h^{-1}$ Gyr. For most rotation axis orientations, including those not aligned with any principal axis, DM only halos have relatively stable shapes, whereas DM + baryon halos rotating about an axis other than the minor or major axis may have their shapes evolve significantly over the duration of the figure rotation.}
    \label{fig:rotation_axis_stability}
\end{figure*}

An exhaustive study of the causes of figure rotation, its evolution, and its stability is beyond the scope of this paper. In this subsection, we instead expand on the evidence we observe in prior subsections which is suggestive of enhanced stability of rotations about the halo minor axis. We observed two pieces of evidence in prior subsections in support of this: the highest density of observed rotation axes are aligned with the halo minor axis (fig \ref{fig:octant_prob}) and longer-lived rotations are preferentially aligned with the halo minor axis (fig \ref{fig:minor_axis_alignment_v_tD}). Both of these effects are greater in the presence of baryons relative to the DM only simulation. 

Steady figure rotation is dependent on the halo shape remaining relatively stable and triaxial over the rotation lifetime by definition. If the axis lengths evolve significantly, then figure rotation may be disrupted. For example, halos are expected to undergo a degree of oscillatory ``ringing'' in their quadrupole moments from past accretion \cite{carlberg_milky_2019}. If the amplitude of this ringing is substantial it could in principle lead to time-evolving halo axis lengths and shape parameters. Further, figure rotation could be completely disrupted if the halo shape evolves to be oblate, prolate, or spherical.

Halo shapes which are changing on relatively short timescales may indicate that orbits necessary to maintain the triaxial figure are destabilized. For instance \cite{deibel_orbital_2011} show that for certain triaxial shapes and pattern speeds, figure rotation can cause an increase in the fraction of chaotic orbits. Since chaotic orbits do not conserve 3 integrals of motion, they can undergo a process called ``chaotic mixing'' \citep{merritt_chaos_1996} that results in individual orbits evolving towards rounder shapes. Such orbit evolution (especially if it involves a moderate fraction of orbits) can result in the evolution of the underlying self-consistent potential to a more oblate or spherical shape and  would simultaneously decrease the ability of our algorithm to detect figure rotation. The process is also likely to convert the angular momentum associated with figure rotation to the angular momentum associated with streaming motion.

To examine the stability of figure rotation as a function of rotation axis orientation, we estimate the timescales on which the halo shapes are expected to evolve significantly. We define this timescale by the halo $p/e$ shape divided by the average time derivative of $p/e$ over the duration of the rotation. Results for the dependence of these timescales on the rotation axis orientation are shown in figure \ref{fig:rotation_axis_stability}. In both DM-only and DM+B simulations, we find that most halos rotating about the minor or major axis have shapes which are stable for greater than one Hubble time, but halos rotating about the intermediate axis have unstable $p/e$ shapes which evolve significantly within a few Gyr. Halos rotating about an axis which does not align with any principal axis typically have stable shapes for $\gtrsim 5 h^{-1}$ Gyr in the DM-only simulation, but have relatively unstable shapes in the DM+B simulation evolving on timescales $\lesssim 5 h^{-1}$ Gyr. This lower shape stability in the DM+B simulation may offer an explanation for the lower prevalence of figure rotation lasting at least $\sim 330 h^{-1}$ \added{Myr} that we observe in this simulation. 

\section{Summary and Discussion}\label{sec:conclusion}

We conduct a study of the influence of baryonic physics on dark halo figure rotation using the TNG50 and TNG50 DM-only simulations from the IllustrisTNG simulation suite. We use a catalog of halos in the mass range of $10^{10} - 10^{13} ~h^{-1} M_\odot$ in the redshift range $0 \leq z \leq 0.35$ (lookback time $0 \leq t_{lb} \leq 2.7 ~h^{-1}$ Gyr). Within this range, we identify 1,577 halos in the DM-only TNG50 run and 1,396 halos in the mixed-phase DM+baryonic TNG50 runs which are free of major mergers (mass ratio $>$ 1:10) and have mass accretion rates below \replaced{$\sim10\%/ 220 ~h$ Myr$^{-1}$}{10\% between two snapshots spaced $\sim220 h^{-1}$ apart}. We outline our key findings below:
\begin{itemize}
    \item We develop a new methodology for detecting figure rotation about an arbitrary axis and for arbitrary duration of rotations. This method effectively combines the best features of the quaternion method and best-fit plane method introduced by \cite{bailin_figure_2004}. See figures \ref{fig:single axis fitting} and \ref{fig:genetic_algorithm} for an overview of their workflow.

    \item Figure rotation of any duration \replaced{ $\gtrsim 330 ~h^{-1}$}{above our minimum detectable duration of $\sim 330 ~h^{-1}$} Myr was around 12$\%$ less common in DM+B halos. The prevalence of figure rotation \replaced{lasting at least $\sim 330 ~h^{-1}$ Myr}{of any duration within our detection limits of $0.33h^{-1} - 2.7h^{-1}$ Gyr} was $94\%$ (1,484/1,577) in the DM-only simulation compared to $81.6\%$ (1,139/1,396) in the full mixed-phase run. For durations lasting $\geq 2 ~h^{-1}$ Gyr,  we detected coherent rotation in $47\%$ of DM-only halos and $36\%$ of DM+B halos. 
    
    \item In both DM-only and DM+B simulations halos are more likely to have their figure rotation axis aligned with either the minor or major axis of the halo, in agreement with past results \citep{bailin_figure_2004,bryan_figure_2007}. Figure rotation axes which do not align with any principal axis are \added{also} common, representing roughly $35\%$ of DM only halos and $33\%$ of DM+B halos. These rotation axes are typically aligned within the plane containing the major and minor axes. Although much less common ($<1$\%) alignment with the intermediate axis is seen in both types of halos. See figure~\ref{fig:octant_prob}.
    
    \item Longer lived rotations show a higher degree of alignment with the halo minor axis. DM+B halos show a uniformly higher alignment with the halo minor axis compared to DM only halos across all durations. The fraction of halo rotation axes within $20^\circ$ of the halo minor axis was $55\%$ in DM+B halos, as compared to $48\%$ in the DM-only halos. See figure \ref{fig:minor_axis_alignment_v_tD}.
    
    \item The median pattern speeds in the DM+B TNG50 run were $14\%$ faster relative to the DM-only median pattern speed. The pattern speeds in both simulations were lognormally distributed, with a median of $0.25~h$ km s$^{-1}$ and width of $0.83 ~h$ km s$^{-1}$ for the DM-only run and median of $0.29~h$ km s$^{-1}$ and width of $0.81 ~h$ km s$^{-1}$ for the DM+B run. Our fit parameters for the DM-only simulation are less than $1\sigma$ different than those found by \cite{bryan_figure_2007} in their DM-only simulations. See figure \ref{fig:pattern_speed_hist}.
    
    \item The upper edge of the pattern speed distributions decreases with increasing duration out to durations of $\sim 1 ~h^{-1}$ Gyr, at which point the distribution becomes stable. This trend appears similar both in the presence and absence of baryons. The upper edge of the detected pattern speeds with durations $\geq 1 ~h^{-1}$ Gyr matches very closely to the upper stability limit for figure rotation found by \cite{deibel_orbital_2011}. See figure \ref{fig:pattern_speeds_vs_duration}.

    \item Figure rotation which is not aligned with either the halo minor or major axis might cause the halo axial ratios to evolve in time. In figure \ref{fig:rotation_axis_stability} we show that halos rotating about other axes on average have $p/e$ shapes which are anticipated to evolve significantly in less than one Hubble time. Studying the exact origin of this relationship, and whether it is causal, is beyond the scope of this paper and will be reserved for future work.
    
\end{itemize}

Although it has been known for over 3 decades that triaxial dark matter halos in $\Lambda$CDM simulations exhibit figure rotation due to tidal torquing and mergers \citep{dubinski_cosmological_1992,bailin_figure_2004,bryan_figure_2007}, the magnitude of the figure rotation was considered to be too low to produce observable effects on the baryonic components of galaxies. This changed due to the realization that despite its small predicted magnitude, the figure rotation of a DM halo can induce  significant, observable effects on stellar tidal streams, especially those streams that extend over a large range of radii like the Sagittarius tidal stream \citep{valluri_detecting_2021}. There are now nearly 100 identified streams in the Milky Way (MW), more than half of which have proper motions from \textit{Gaia} \citep{mateu_galstreams_2022} with the potential to measure radial velocities using spectroscopic surveys like DESI \citep{desi_collaboration_desi_2016, prieto_preliminary_2020, cooper_overview_2022}. The future Roman Space Telescope will have the ability to retrieve proper motions of halo stars down to $\sim$ 10 $\mu$as per year with a depth of 25th magnitude in the filter centered at 1.46 micron \footnote{\url{https://www.stsci.edu/files/live/sites/www/files/home/roman/_documents/roman-capabilities-stars.pdf}}. This will enable the detection of streams to much greater galactocentric radii than is possible with \textit{Gaia}, in addition to much greater completeness of known streams. The availability of high quality stream data is likely to continue growing in the coming years, and constraining figure rotation in the MW using tidal streams is becoming increasingly feasible. Simultaneously, many streams are also being detected around external galaxies with many more expected from HSC, LSST, Roman (formerly $\textit{WFIRST}$), and $\textit{Euclid}$ \citep{pearson_detecting_2019}.

All previous studies of figure rotation were restricted to DM-only cosmological simulations. Since those works in the mid 2000s it has been learned that baryons significantly modify the shapes and central concentrations of dark matter halos \citep{kazantzidis_effect_2004, zemp_impact_2012,chua_shape_2019, prada_dark_2019}. This motivated us to revisit the issue of figure rotation in the contemporary TNG50 suite of cosmological hydrodynamical simulations as well as their DM-only analogs. This study clearly demonstrates that the presence of baryons does not eliminate the ability of dark matter halos to exhibit figure rotation, rather it might cause them to rotate slightly more rapidly. 

This robustness of figure rotation to the presence of baryons in $\Lambda$CDM is important since figure rotation can only arise from a triaxial matter distribution composed of dark matter particles and cannot arise in alternative gravity theories such as MOND \citep{bailin_figure_2004}.
In addition to the well known $\Lambda$CDM paradigm, dark sector models include warm dark matter (WDM) \citep{bond_massive_1980, avila-reese_formation_2001,bose_reionization_2016} and fuzzy (or ultra-light bosonic) dark matter (FDM) \citep{hui_ultralight_2017}. Other theories consider differences in the  dynamical behavior of the dark matter particle (e.g. superfluid DM, \citealt{berezhiani_phenomenological_2018}), or its interaction strength (e.g. self-interacting dark matter (SIDM), \citealt{spergel_observational_2000,  tulin_dark_2018}).
Future studies of figure rotation  of halos in DM-only and DM+baryonic simulations with SIDM, WDM, FDM etc. should be carried out to assess whether the halos in these alternative scenarios also exhibit figure rotation and if their pattern speeds, durations and rotation axis orientations are distinguishable from $\Lambda$CDM. This could provide new avenues for studying the nature of the dark matter particle and distinguishing particle theories from theories like MOND.

An important result of our study is that the figure rotation axis in a significant fraction of galaxies with and without baryons is misaligned with the principal axes of the halos. In future it would be instructive to examine the alignment/misalignment of disk angular momentum with halo rotation axis since any misalignments could lead to detectable warps especially in extended HI disks.

Our study of TNG50 halos shows that individual halos can experience multiple epochs of figure rotation with different pattern speeds and rotation axes. A thorough investigation of the causes of these multiple epochs of rotation, and their impact on the structure of the DM and baryonic halo components, is warranted. In particular, the behavior of disky baryonic components in response to these changes in the halo figure rotation could lead to interesting observable signatures for figure rotation in distant halos. Such a study will be carried out in a future paper.

Our study used a highly restricted sample of halos since the standard shape tensor method used to determine halo shapes is extremely sensitive to subhalos whose orbits can result in spurious figure rotation signatures. However, in recent years it has become increasingly clear that the influence of the Large Magellanic Cloud (LMC), now believed to be between 1/6th and 1/4th the mass of the Milky way, is significantly perturbing the MW halo. These perturbations in the gravitational potential of the dark matter halo are thought to be both time-evolving and radius dependent \citep{garavito-camargo_quantifying_2021}, and have been shown to affect the structure and kinematics of several tidal streams \citep{erkal_total_2019, vasiliev_tango_2021,koposov_s_2023}. 

It is possible or even likely that the influence of the LMC is consequential for figure rotation in the MW dark halo. Depending on the initial orientation of the MW dark halo, the gravitational influence of the LMC could possibly torque the halo, which may in turn either enhance or disrupt figure rotation. However, due to the sensitivity of the shape tensor method to the presence of massive satellites, this study was conducted with a restrictive set of merger-free halos which are not representative of the MW-LMC system, though the MW provides the most promising venue for measuring halo figure rotation. Studying figure rotation in the presence of nearby massive companions will require the development of novel methods which are insensitive to halo substructure and is beyond the scope of this paper. We will explore the subject of angular momentum transfer and figure rotation in the presence of massive, LMC-like companions for future work.

\begin{acknowledgments}
The authors would like to thank the IllustrisTNG collaboration for generously providing access to TNG50 and TNG50\_DM simulation data and computational resources via the online jupyterlab workspace. The IllustrisTNG simulations were undertaken with compute time awarded by the Gauss Centre for Supercomputing (GCS) under GCS Large-Scale Projects GCS-ILLU and GCS-DWAR on the GCS share of the supercomputer Hazel Hen at the High Performance Computing Center Stuttgart (HLRS), as well as on the machines of the Max Planck Computing and Data Facility (MPCDF) in Garching, Germany. The authors gratefully acknowledge support from NASA-ATP award 80NSSC20K0509. The authors thank Eric Bell and Leandro Beraldo e Silva for comments and discussions that improved this paper. MV is pleased to acknowledge UM undergraduate student Joseph Hofer whose initial efforts on determining the figure rotation of 10 TNG-100 halos motivated this study. MV \& NFA are gratefully acknowledge financial support from NASA-ATP award  80NSSC20K0509 to MV.

\end{acknowledgments}

%

\vspace{5mm}


\software{astropy \citep{the_astropy_collaboration_astropy_2018},  
AGAMA \citep{vasiliev_agama_2019}
numpy \citep{harris_array_2020}, 
scipy \citep{virtanen_scipy_2020}.        
          }



\appendix

\section{Pattern speed independence of halo mass}\label{appendix:mass_pattern_speed}

\begin{figure}
    \centering
    \includegraphics[width=0.4\textwidth]{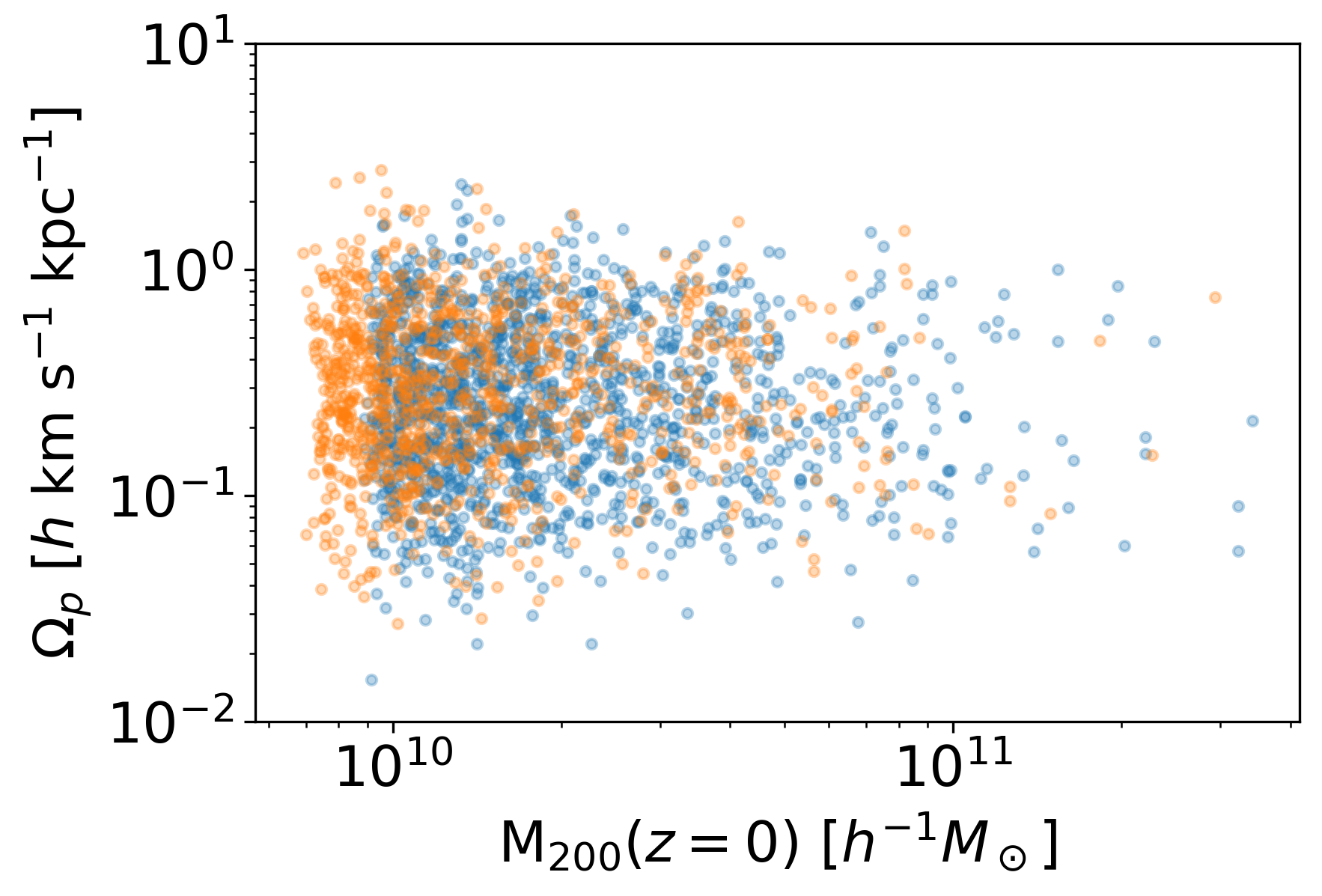}
    \includegraphics[width=0.4\textwidth]{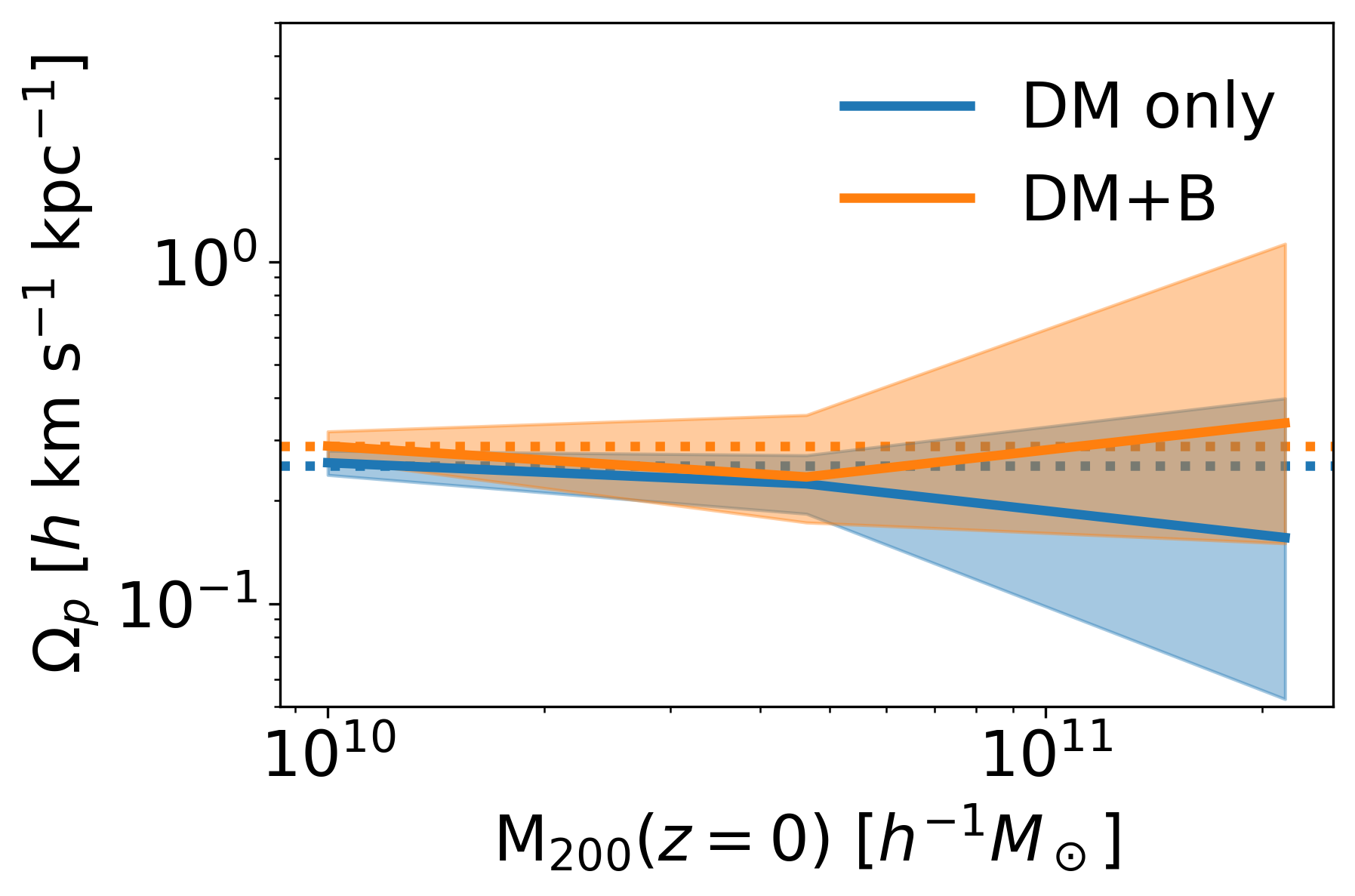}
    \caption{Observed figure rotation pattern speeds are independent of the halo virial mass. \textbf{Left:} Scatter plot of measured pattern speeds against the halo virial mass at $z=0$ for both DM only (blue) and DM+B (orange) runs of TNG50. \textbf{Right:} Medians of lognormal distributions fit in 4 different mass bins spanning $10^{10}h^{-1} - 10^{12}h^{-1}$ M$_\odot$ using bootstrap fitting with 500 samples per mass bin. The shaded region shows the upper and lower bounds of these bootstrap samples. Dashed lines indicate the full population medians. Results of the bootstrap analysis are consistent with pattern speeds which are independent of halo mass.}
    \label{fig:no_mass_dependence}
\end{figure}

\cite{bailin_figure_2004} found that figure rotation is independent of halo mass. Our halo catalog allows us to study whether this independence remains true for lower halo masses. To assess any potential mass dependence, we split our detected pattern speeds into 4 mass bins spanning the $10^{10}h^{-1} - 10^{12} h^{-1}$ M$_\odot$ virial mass range containing the halos within our merger-free catalog (within TNG50, we found no halos between $10^{12}h^{-1}-10^{13}h^{-1}$ M$_\odot$ satisfying the criteria outlined in section \ref{sec:halo_catalog}). In each of the 4 mass bins, we fit the observed pattern speeds to a lognormal distribution using the same bootstrap routine as described in section \ref{sec:pattern_speed_dist}. The bottom panel of figure \ref{fig:no_mass_dependence} shows the median bootstrap fit results as well as the upper and lower bounds of the bootstrap fits. The results we obtain from this analysis are consistent with figure rotation pattern speeds which are independent of halo mass.


\bibliography{references}{}
\bibliographystyle{aasjournal}


\listofchanges
\end{document}